\documentclass[10pt,a4paper]{article}
\usepackage{graphicx}
\usepackage{amsmath}
\usepackage{url}
\usepackage{bm}
\usepackage{longtable}
\usepackage{adjustbox}
\usepackage{multirow}
\usepackage[subrefformat=parens]{subcaption}
\usepackage{hyperref}
\usepackage{authblk}
\usepackage{array}

\newcommand\figref[1]{Fig.~\ref{#1}}
\newcommand\tabref[1]{Table~\ref{#1}}

\begin{document}

\renewcommand{\thefootnote}{\fnsymbol{footnote}}  

\title{Untangling the complexity of market competition in consumer goods\\-- A complex Hilbert PCA analysis}

\renewcommand\Authands{, }
\renewcommand\Affilfont{\small}

\author[1]{Makoto Mizuno$^\dagger$}
\author[2,3]{Hideaki Aoyama}
\author[4]{Yoshi Fujiwara}

\affil[1]{School of Commerce, Meiji University, Tokyo 101-8301, Japan}
\affil[2]{Research Institute of Economy, Trade and Industry, Tokyo 100-0013, Japan}
\affil[3]{RIKEN iTHEMS, Wako, Saitama 351-0198, Japan}
\affil[4]{Graduate School of Simulation Studies, University of Hyogo, Kobe 650-0047, Japan}

\footnotetext[2]{Corresponding author: \texttt{makmizuno@gmai.com}}

\date{August 24, 2020}

\maketitle

\begin{abstract}

Today's consumer goods markets are rapidly evolving with significant growth in the number of information media as well as the number of competitive products. In this environment, obtaining a quantitative grasp of heterogeneous interactions of firms and customers, which have attracted interest of management scientists and economists, requires the analysis of extremely high-dimensional data. Existing approaches in quantitative research could not handle such data without any reliable prior knowledge nor strong assumptions. Alternatively, we propose a novel method called complex Hilbert principal component analysis (CHPCA) and construct a synchronization network using Hodge decomposition. CHPCA enables us to extract significant comovements with a time lead/delay in the data, and Hodge decomposition is useful for identifying the time-structure of correlations. We apply this method to the Japanese beer market data and reveal comovement of variables related to the consumer choice process across multiple products. Furthermore, we find remarkable customer heterogeneity by calculating the coordinates of each customer in the space derived from the results of CHPCA. Lastly, we discuss the policy and managerial implications, limitations, and further development of the proposed method.

\end{abstract}

\bigskip
\begin{flushleft}
\end{flushleft}

\renewcommand{\thefootnote}{\arabic{footnote}}
\setcounter{footnote}{0} 

\clearpage
\section{Introduction}

In rapidly evolving, highly competitive consumer goods markets, firms are faced with a number of factors affecting business. These factors often veer away from conventional rules of thumb or theory and can be mutually interacting, providing the challenge of detecting the meaningful relationships between such factors without any strong assumptions. This challenge represents the complexity of consumer choice process caused by the proliferation of viable marketing instruments and competing products. Traditionally, consumer purchase could have been a function of price, radio or TV advertising. Recently, with the penetration of the Internet and mobile devices the possible set of marketing instruments have been expanding. Some consumers might search for information using mobile phones while others might gain detailed information by visiting related websites after watching TV advertisements. If the strategy for mixing these instruments is heterogeneous across competitive firms or products, the increased number of competitors would intensify the resulting complexity further.

To grasp the time sequence of exposures to multiple marketing instruments, the concept of 'customer journey map' is gaining popularity with practitioners and researchers \cite{edelman2015}  \cite{lemon2016}, for which some quantitative models have been proposed \cite{anderl2016} \cite{dehaan2015}. This approach focuses on the consumer decision process for a single product/firm only, ignoring competition among multiple products. Hence, it fails to reveal real consumer behaviors for the market with multiple competitive products. In order to deal with dynamic market competition across products, multivariate time-series analyses are highly popular; and have been developed in econometrics and intensively applied to assess the long-term effectiveness of marketing instruments such as pricing or advertising. This method includes a variety of models such as the vector auto-regressive model \cite{pauwels2007}, dynamic linear model \cite{ataman2010}, varying parameter model \cite{sriram2007}, and the Kalman filter \cite{kolsarici2016}. These models are evaluated in terms of whether they satisfy the following conditions \cite{ataman2010}:

\begin{itemize}
\item Endogeneity: The recent quantitative marketing studies inspired by economics has increasingly treated variables of marketing instruments as endogenous variables, not as exogenous variables freely determined by firms. Suppose customers who are more likely to purchase a focal product are targeted in an advertising campaign for the product. In this case, observed advertising data would reflect the potential purchases of the product. With no consideration for this endogeneity, the effects of advertising on purchases would be overestimated.  
\item Performance feedback: Even if firms do not foresee future performance, their marketing actions would be constrained by past performance due to performance-based budget allocation. Hence the lagged effect of the performance to the activity level of marketing instruments could exist.
\item Competitive reactions: Firms that compete in the same market, are sensitive to competitors' actions, often causing retaliation. This has been studied as a hot issue in marketing science, using a variety of data and analytical methods \cite{steenkamp2005}. 
\end{itemize}

Even multivariate time-series analyses often face serious difficulties in handling the complexity of current consumer markets, because of the limited number of tractable parameters. Although most consumer goods markets are composed of many more firms/products and marketing instruments, the above extant models have only been applied to cases with only three to five competitors and a few marketing instruments. The competition analyzed in this study consists of 18 products and 4 marketing instruments (price, TV advertising, search via mobile devices or PCs, and web visits via mobile devices or PCs). Applying the conventional time-series analyses for these markets would cause an explosive increase in possible parameters to be estimated. To avoid this, researchers have relied on their experience or research tradition and have been forced to impose strong restrictions at the risk of losing crucial information on the complexity of the focal markets. 

Alternatively, we propose a novel method called complex Hilbert principal component analysis (CHPCA) \cite{aoyama2010econophysics,aoyamacambridge2017} to unravel the complexity of the consumer choice process across multiple competitive products. CHPCA was developed originally in econophysics as an extension of Principal Component analysis (PCA) to uncover temporal comovements among variables observed in the macro economy \cite{kichikawa2020ilp} or foreign exchange markets \cite{vodenska2016}. CHPCA can handle massive high-dimensional time-series data without any strong assumptions about the phenomenon of interest. In addition, this method satisfies the above-mentioned three conditions for evaluating marketing models with multiple competitive products \cite{ataman2010}: CHPCA satisfies the first condition (endogeneity) since it treats all variables (marketing actions by instrument and performance indexes) as comoving variables. Additionally, this method satisfies the second condition (feedback from past outcomes) and the third condition (competitive reactions) since these are incorporated as part of possible comovements.

Another advantage of CHPCA is its practicality. Practitioners can solve real problems under time and effort constraints. First, as with ordinal PCA, users of CHPCA may not need any prior knowledge or assumptions regarding the phenomenon. By using random rotation simulation (RRS), significant eigenmodes (principal components in PCA) can be selected automatically in a theoretically justifiable manner. Second, the results of CHPCA are easily interpreted on the complex plane corresponding to each eigenmode following stylized procedures. Conveniently, the information obtained is integrated and visualized by a synchronized network with Hodge decomposition. Finally, this method provides a skeleton of consumer choice process across competitive products using aggregate marketing data, which is widely available for the packaged consumer goods markets. The model also provides information on heterogeneity in customer profile.

We recommend CHPCA for conducting exploratory analyses. Similar to the division of roles between exploratory and confirmatory factor analyses, CHPCA can be complementary to traditional multivariate time-series analyses. CHPCA is used to generate hypotheses on possible causalities among a huge number of variables while multivariate time-series analyses are used to rigorously test hypotheses. If a few critical relationships are detected as a result of the analysis using CHPCA, we can apply quasi-experimental methods such as a regression discontinuity design \cite{hartman2011} or a propensity score method \cite{mizuno2006}, which have been used to prove causality in observational data. 

The remainder of this paper is organized as follows: In section 2, we describe the data for the beer market in Japan. We propose the application of CHPCA to understand the consumer choice process across multiple products in section 3 and the procedure for depicting the skeleton of the consumer choice process via a synchronization network in section 4. The procedure to detect customer heterogeneity and the results are reported in section 5. In section 6, we conclude our paper with a discussion on the policy and managerial implications, the remaining problems, and avenues for further research.

\section{Data}\label{sec:data}
\subsection{Data Collection}

In this study, we analyze the comovement of consumer purchases (quantities and prices paid) of beer and related marketing communication activities. The reason for our focus on this market is that it is a typical monopolistic-competitive market, where a few firms compete with differentiated products using a full range of marketing instruments such as TV advertising, web/mobile marketing, price promotion, etc., attracting attentions of economic policymakers and marketers. Hence, we use INTAGE Single-source Panel (i-SSP) data, which is the most comprehensive consumer database commercially operated in Japan measuring daily purchases of a wide variety of consumer package goods and consumer marketing communication activities (exposure to TV ads, visits to web sites via mobile device/PC, and search activities via mobile device/PC. For the current analysis, we use these data for 365 days from April 1, 2013 to March 31, 2014 (inclusive). The abbreviation and description of each time series is shown in \tabref{tab:abbs}. 

\begin{table}
        \centering
        \caption{\bf Abbreviation and the description of the five kinds of time-series}
        \scalebox{0.9}{
        \begin{tabular}{c|l}
        Abbreviation&Description\\
        \hline
        P&Price per unit quantity (yen/m$\ell$)\\
        Q&Quantity purchased (m$\ell$)\\
        Visit&Visit to related web sites via mobile device or PC (seconds) \\ 
        TVAd&Exposure to TV advertising (seconds) \\
        Search&Search frequency via mobile device or PC (times) \\
        \hline
        \end{tabular}}
        \label{tab:abbs}
\end{table}
    
These data initially capture individual-level behaviors of the panel. 
For this analysis, however, we aggregated the data over all customers due to the limited size of the data. The potential heterogeneity among customers is represented as the location of a few-dimensional space (see the Customer Profile Section.
The purchase data are documented at the store-keeping unit (SKU) level
while the advertising and other communication variables are documented at the product level. In general, one product is composed of multiple SKUs. When merging the two types of data, the purchase data are summed across SKUs by their corresponding product. 

\begin{figure}
    \centering
    \includegraphics[width=0.7\textwidth]{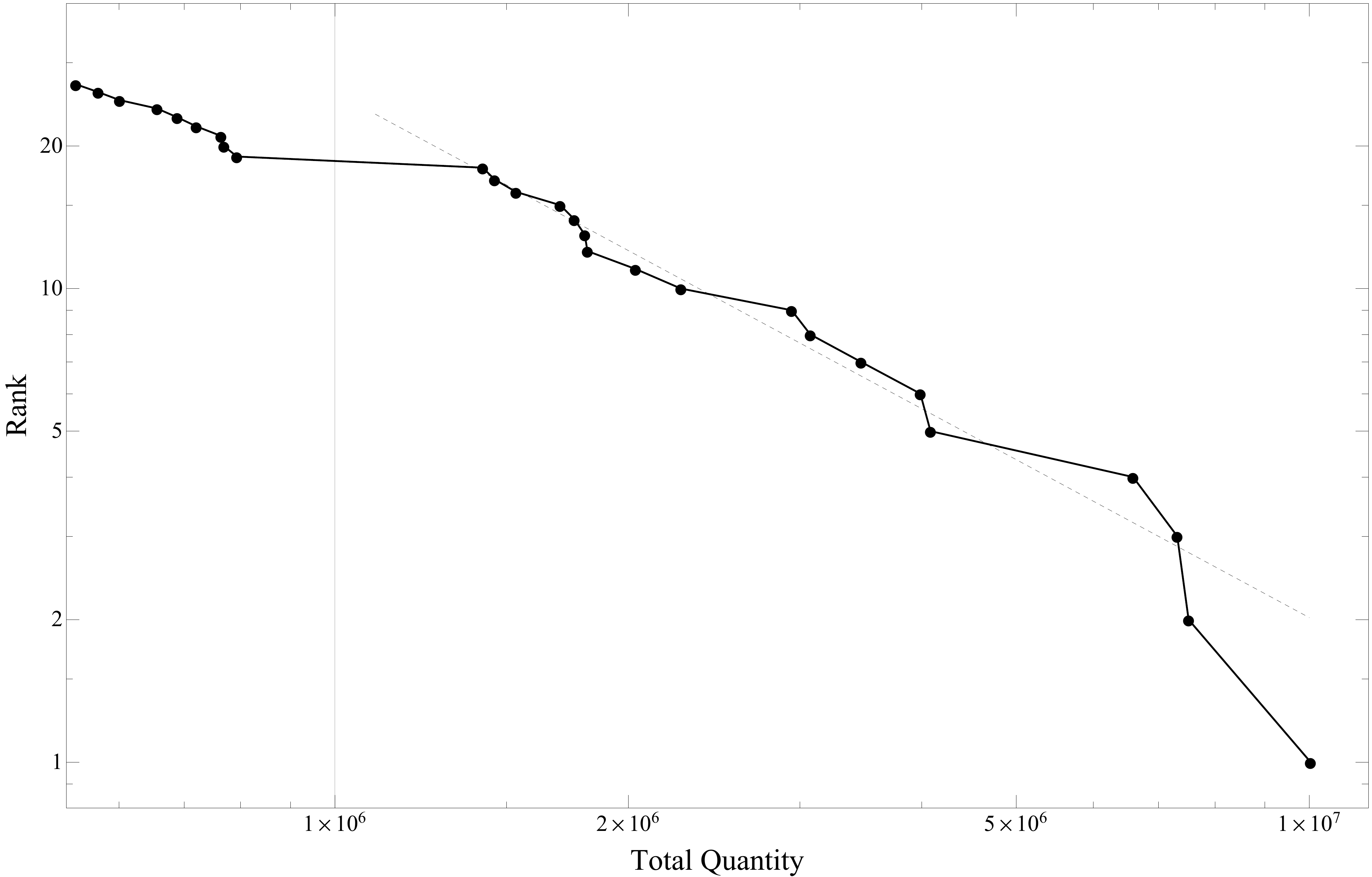}
    \caption{\bf Rank-size plot of the top selling products.}
    \label{fig:top18}
\end{figure}

There are 163 beer products from 14 firms in the original data. We selected the top 18 products according to the total quantity in the data during the stated period: \figref{fig:top18} is the rank-size plot of some of the top products. The top 18 products selected for the current analysis are those beyond the thin vertical line, Total Quantity $>1\times10^4$.
As is apparent in this plot, these products form a distinctive top group with a large gap 
between this group and the followers. Furthermore, this top group roughly obeys the power-law indicated by
the thin dashed line, with [Rank] $\propto$ [Total Quantity]$^{-1.109}$.

\begin{figure}
    \centering
    \includegraphics[width=0.7\textwidth]{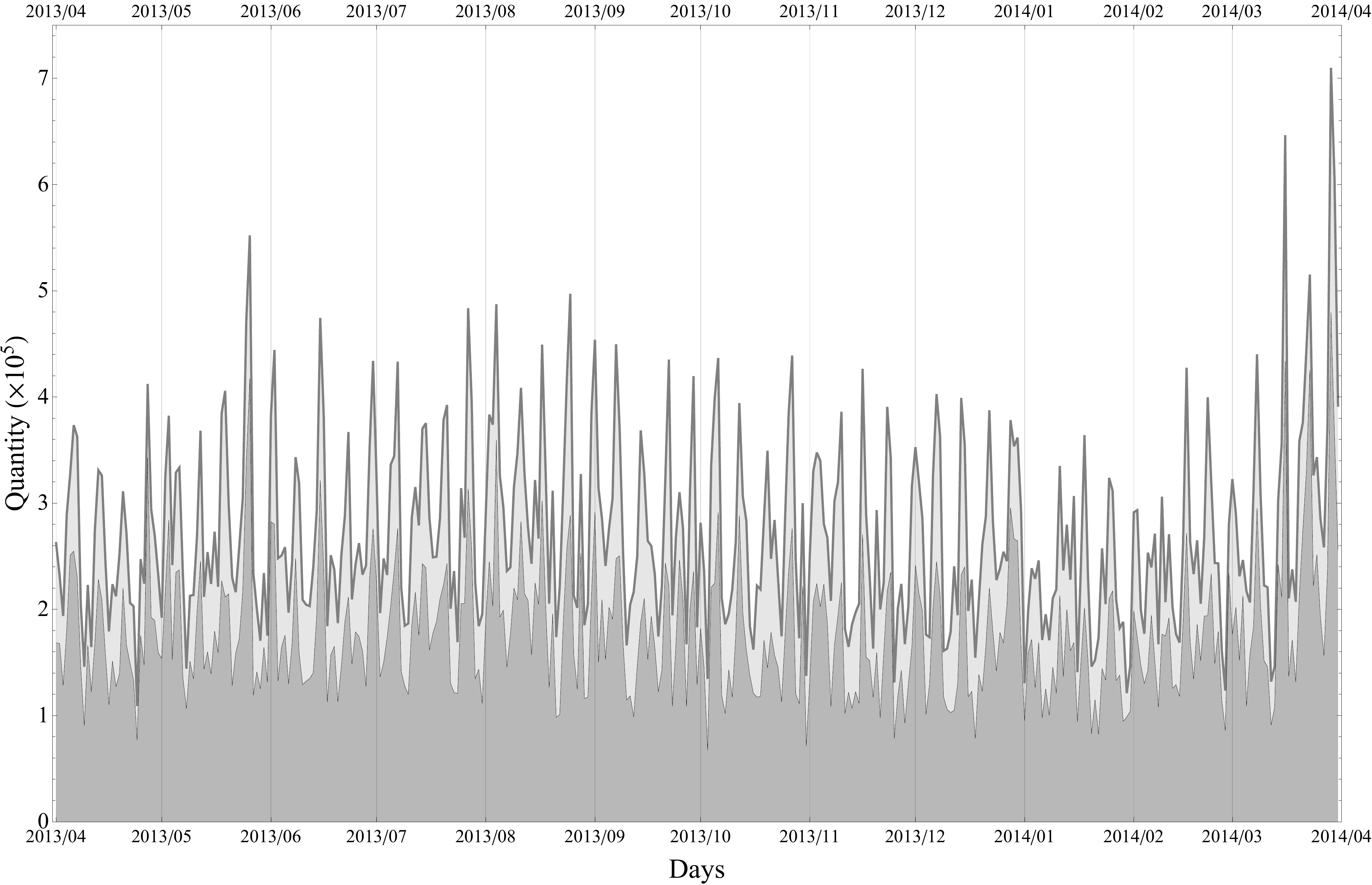}
    \caption{\bf Daily total quantities. Light gray: all products, Dark gray: Selected 18 products.
    Apparent periodic peaks correspond to Sundays.}
    \label{fig:Q}
\end{figure}

The data for the top 18 products cover 64.9\% of all the sales.
\figref{fig:Q} shows the daily total quantities for both all products (in light gray) and
the selected 18 products (in dark gray). Periodic peaks apparent in this plot occur at weekends when quantity rises on Saturdays and peaks on Sundays.
The high peak structure at the end of the period; {\it that is,} at the end of March 2014, is explained by the VAT hike from 5\% to 8\% on April 1, 2014.\footnote{%
We have data beyond this day for another several months. However, we decided to take this one-year period to avoid bias from seasonal dependence and the strong influence from the increase in quantity (and the downfall on and after April 1, 2014).}

The products are shown with codes, the first letter of which corresponds to the firm, and the second letter (digit) corresponds to the product:
For example, the code ``A1'' means that the product is from firm `A' and the product `1'.
With five types of data for each product as listed in \tabref{tab:abbs},
we have $18\times5=90$ time series altogether.
However as no communication activity was observed for some products, 
we set the threshold to 51 days: we used only the time-series data with
51 or more days of entry (more than or equal to once a week).
With this threshold, we have 65 time series for purchases and related communication activities, which are listed in \tabref{tab:days}.

\newcolumntype{R}{>{\raggedleft\arraybackslash}p{2.3em}}
\newcolumntype{C}{>{\centering\arraybackslash}p{2.3em}}
\begin{table}
        \centering
     \caption{\bf Top-selling 18 products and availability of data in days.}
     \scalebox{0.85}{
\begin{tabular}{c|c|r|RRRRR}
\hline
 \text{Rank} & \text{Code} & \text{Total Sales} & \text{P} & \text{Q} &
   \text{Visit} & \text{TVAd} & \text{Search} \\
   \hline
 1 & \text{D2} & $1.00\times 10^7$ & 365 & 365 & 158 & 328 & 209 \\
 2 & \text{A4} & $7.52\times 10^6$ & 365 & 365 & 25 & 287 & 163 \\
 3 & \text{B1} & $7.32\times 10^6$ & 365 & 365 & 164 & 306 & 153 \\
 4 & \text{C3} & $6.59\times 10^6$ & 365 & 365 & 29 & 173 & 76 \\
 5 & \text{B3} & $4.09\times 10^6$ & 354 & 354 & 363 & 354 & 73 \\
 6 & \text{A3} & $3.99\times 10^6$ & 357 & 357 & 29 & 249 & 34 \\
 7 & \text{A1} & $3.47\times 10^6$ & 360 & 360 & 63 & 185 & 117 \\
 8 & \text{B5} & $3.08\times 10^6$ & 365 & 365 & 62 & 274 & 19 \\
 9 & \text{D3} & $2.94\times 10^6$ & 339 & 339 & 3 & 0 & 0 \\
 10 & \text{D1} & $2.27\times 10^6$ & 355 & 355 & 284 & 334 & 270 \\
 11 & \text{B4} & $2.03\times 10^6$ & 258 & 258 & 0 & 0 & 0 \\
 12 & \text{A6} & $1.82\times 10^6$ & 330 & 330 & 0 & 0 & 0 \\
 13 & \text{B2} & $1.80\times 10^6$ & 298 & 298 & 172 & 228 & 41 \\
 14 & \text{C1} & $1.76\times 10^6$ & 341 & 341 & 82 & 219 & 168 \\
 15 & \text{A2} & $1.70\times 10^6$ & 304 & 304 & 5 & 111 & 5 \\
 16 & \text{C2} & $1.53\times 10^6$ & 342 & 342 & 39 & 216 & 25 \\
 17 & \text{A5} & $1.46\times 10^6$ & 276 & 276 & 0 & 0 & 0 \\
 18 & \text{C4} & $1.42\times 10^6$ & 198 & 198 & 0 & 0 & 0 \\
 \hline
   \end{tabular}}
    \label{tab:days}
\end{table}

\clearpage

\subsection{Descriptive Statistics}

\tabref{tab:ynav} summarizes the means and standard deviations for the time series.
Symbol "---" implies too sparse data due to no communication activity.
We observe that these time series are highly volatile in temporal change. 
We then use the standard method of subtracting the mean and dividing it by
the standard deviation. Let us denote the resulting time series by
$x_\alpha(t_i)$ where $\alpha=1,\cdots,N(=65)$ is the label for the time series,
and $t_i=1,\cdots,365$ denotes the number of days. The mean and standard deviation of $x_\alpha(t_i)$ are 0 and 1 respectively,
for which we apply our methodology explained in the next section.

\begin{table}
        \centering
        \caption{{\bf Descriptive statistics for the top-selling 18 products} Means and standard deviations (in parentheses).
    The symbol "---" corresponds to discarded data because of sparsity (see text).}
    \scalebox{0.8}{
    \begin{tabular}{c|C|rrrrr}
 \text{Rank} & Code &\text{P} & \text{Q} & \text{Visit} & \text{TVAd} & \text{Search} \\
 \hline
  1 & D2 & 0.287(0.011) & 27483.0(14758.6) & 18.9(66.6) & 4770.1(6042.2) & 1.92(3.59) \\
  2 & A4 & 0.294(0.014) & 20596.9(14640.2) & --- & 4164.6(6126.4) & 1.30(3.07) \\
  3 & B1 & 0.494(0.026) & 20052.2(13850.3) & 75.1(291.5) & 4004.4(6704.8) & 2.15(5.15) \\
  4 & C3 & 0.287(0.015) & 18064.9(13970.4) & --- & 1271.7(3159.3) & 0.54(1.76) \\
  5 & B3 & 0.295(0.023) & 11192.3(10534.8) & 175.6(365.3) & 4792.1(5725.9) & 0.50(1.92) \\
  6 & A3 & 0.350(0.025) & 10926.4(11359.7) & --- & 2964.1(4648.4) & --- \\
  7 & A1 & 0.506(0.035) & 9499.7(8786.6) & 9.5(38.5) & 4232.1(7763.3) & 1.10(2.56) \\
  8 & B5 & 0.301(0.023) & 8424.9(8117.4) & 19.3(118.3) & 2861.0(3955.7) & --- \\
  9 & D3 & 0.285(0.020) & 8060.4(8510.2) & --- & --- & --- \\
 10 & D1 & 0.567(0.039) & 6205.9(5934.8) & 73.7(184.8) & 10051.1(10428.0) & 3.81(6.15) \\
 11 & B4 & 0.296(0.018) & 5573.2(8665.4) & --- & --- & --- \\
 12 & A6 & 0.304(0.024) & 4975.0(6037.5) & --- & --- & --- \\
 13 & B2 & 0.362(0.028) & 4943.8(6515.0) & 46.2(115.2) & 1769.2(3712.4) & --- \\
 14 & C1 & 0.555(0.046) & 4823.9(5755.2) & 10.6(42.9) & 2486.8(4732.9) & 1.46(3.41) \\
 15 & A2 & 0.362(0.039) & 4665.9(6790.5) & --- & 1359.0(4166.9) & --- \\
 16 & C2 & 0.508(0.041) & 4202.3(5325.6) & --- & 2019.6(5922.4) & --- \\
 17 & A5 & 0.298(0.021) & 3995.4(6606.3) & --- & --- & --- \\
 18 & C4 & 0.296(0.028) & 3884.6(6819.0) & --- & --- & --- \\
   \end{tabular}}
    \label{tab:ynav}
\end{table}

\begin{figure}
  \centering
  \includegraphics[width=0.7\textwidth]{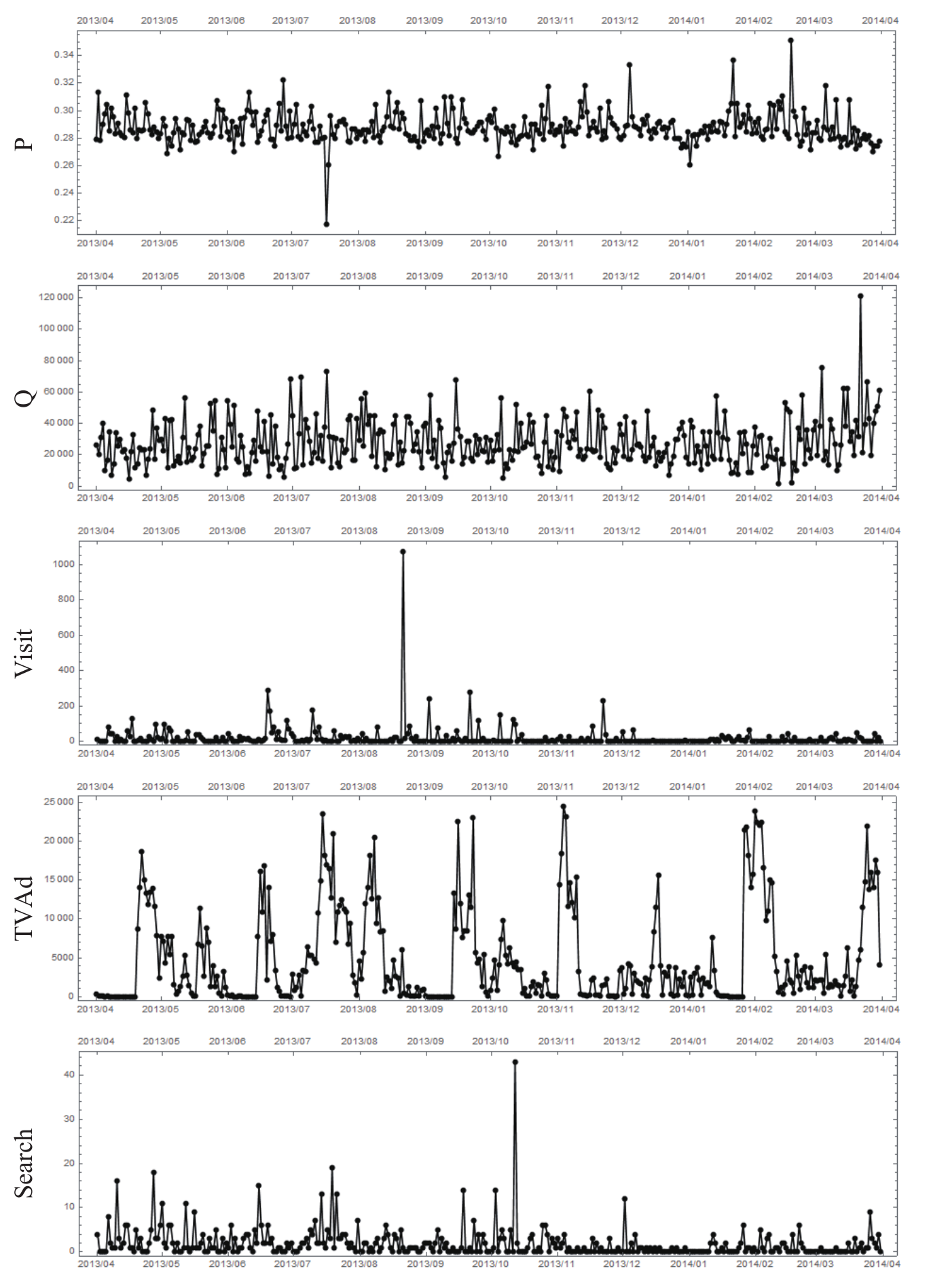}
  \caption{{\bf Sample of time-series} $x_i(t)$ for the top product ``D2''
    for the five kinds of time-series, namely,
    P, Q, Visit, TVAd, and search from top to bottom.}
  \label{fig:ts_samp}
\end{figure}

We depict, as a sample of $x_\alpha(t_i)$, the five types of time series for the top product ``D2" ($\alpha=1,\cdots, 5$) in \figref{fig:ts_samp}.
Price per unit quantity (P) is mostly stable but has spikes of increasing or decreasing price change.
Quantity (Q) has volatility due to growing and sluggish sales.
The frequency of site visits via PC or mobile device (Visit) have tranquility with sudden and short activities. The frequency of TV advertising exposure (TVAd) has weak periodic behavior presumably due to the TV advertising activities of the product's firm and corresponding exposure to customers. The frequency of searches via PC or mobile device (Search) have continuous activities with bursts. 

\section{Complex Hilbert Principal Component Analysis (CHPCA)}\label{sec:method}
\subsection{Method}

Any set of real world time-series data contains information on the behavior of individual time series and the inter correlations in the time series.
In this study, we are interested in inter correlations in the time series.
To identify the structure and 
dynamics of the customer choice process, we 
extract information on the inter correlation between price, sales, and other media approaches.
Such comovement in the data set involves time lead and delay.
Some time series follow other time series because of direct and indirect causal relationships.
Here, our aim is to set up a methodology suitable for detecting inter relationships with time delay.

Principal component analysis (PCA) fulfills our goal partially. For this method,
we calculate correlations between time series and identify the eigenmodes of the
correlation matrix, which are independent comovements in the system.
The larger the eigenvalue, the more significant the presence of the eigenmode.
Some of the eigenmodes, however, are simply the result of random movements in the system.
To identify which modes are significant real comovements, people
often apply random matrix theory, which predicts the eigenvalues from the random time series.
This method has several shortcomings.
\begin{enumerate}
    \item[(i)]
    When seeking comovements with time lead/delay, the time series is shifted relative to other time series to maximize the absolute value of the correlation coefficient. This is feasible for two time series but not so for    a large number of time series. With 100 time series, for example, pair wise calculation is required for nearly 5,000 pairs.
    Then, there is the problem of combining them to obtain system-wide comovements.
    \item[(ii)]
    Random matrix theory (RMT) is practical on that the length of the time series ($T$) and the
    number of the time series ($N$) are both infinite with their ratio ($T/N$) kept
    finite, and all the time series has trivial auto correlation,
    none of which may be satisfied by the real data.   
\end{enumerate}

To overcome these difficulties, we use CHPCA and rotational random simulation RRS. 

The former was originally introduced in
\cite{RACJ1981,Barnett1983,Horel1984,PhysRevLett.107.128501,JOC:JOC1499}
using the Hilbert transformation developed in 
\cite{hilbert,Gabor1946,granger1964spectral,bendat2011random,feldman2011hilbert}
among others.
The approach has been successfully applied in several areas of natural science and economics 
\cite{hilbertikeda,ikeda2013direct,kichikawa2020ilp,vodenska2016}.
We further introduced improvements on CHPCA by
\cite{aoyamacambridge2017}.

In CHPCA, we complexify each of the time series'
Hilbert transformation as an imaginary part and then calculate the complex correlation
matrix. We provide a
pedagogical explanation of the merits of this method.
The Hilbert transformation, simply put, transforms each of the Fourier
components in the manner 
$\cos\omega t \rightarrow -\sin \omega t$ 
and 
$\sin \omega t \rightarrow \cos \omega t$.
Therefore, the complexification converts $\cos \omega t$ 
to $e^{-i\omega t}$
and $\sin \omega t$ to $i\, e^{-i\omega t}$; clockwise rotation on its complex plane.
Furthermore, the Hilbert transformation converts
\begin{equation}
    \cos \omega (t + t_0)=
    \cos \omega t \cos \omega t_0 - \sin \omega t \sin t_0
    \rightarrow
    e^{-i \omega (t + t_0)}.
    \label{eq:iot}
\end{equation}
We denote the complex time series obtained from $x_\alpha(t)$ and standarized (so that its means is equal to zero and its standard deviation is equal to one) as $z_\alpha(t)$.
Complex correlation coefficients (CCC) are defined as inner products of one (complex and normalized) time series ($z_\alpha (t)$) with another;\footnote{Hereafter, $\cdot^*$ denotes the complex conjugate of $\cdot$.}
\begin{equation}
    C_{\alpha\beta}:=\sum_{t} z_\alpha(t) z_\beta^* (t).
\end{equation}
If the time series $\alpha$ and $\beta$ are made of Fourier components of the same $\omega$
but with time constants $t_\alpha$ and $t_\beta$,
the CCC has a phase factor proportional to $t_\alpha-t_\beta$,
the time-difference between the two time series.
If the time series contain multiple Fourier components, the phase of the CCC gives
a nonlinear mean of the time differences of each combination of the
Fourier components. Thus, analysis of the resulting complex correlation matrix
enables us to obtain a view of comovements with system time-lag.
By definition, this is one calculation that avoids any pairwise optimization analysis required by PCA. The eigenmode $\mathbf{e}_n$ of the complex correlation matrix $\mathbf{C}=\{C_{\alpha\beta}\}$ is defined by the following:
\begin{equation}
    \mathbf{C}\mathbf{e}_n=\lambda_n\mathbf{e}_n,
\end{equation}
where the subscript $n$ is defined as the eigenvalues $\lambda_n$ in descending order,
$\lambda_1 \ge \lambda_2 \ge \cdots \ge \lambda_N$.
The eigenvalues satisfy an identity
\begin{equation}
    \sum_{n=1}^N \lambda_n = N.
\end{equation}
The time series are expanded in terms of the eigenmodes:
\begin{equation}
    \mathbf{x}(t)=\sum_{n=1}^N s_{n}(t) \mathbf{e}_n,
\end{equation}
where the coefficient $s_{n}(t)$ is called the \textit{mode signal}, satisfying 
\begin{equation}
\lambda_n=\sum_{t=1}^T |s_n(t)|^2.    
\label{eq:eigenstrength}
\end{equation}
In this sense, the eigenvalue $\lambda_n$ is the
strength of the presence of the corresponding
eigenmode $\mathbf{e}_n$.

To avoid using the RMT, we employ RRS, introduced by \cite{iyetomiarai2013}.
This is done by 
(1) ```rotating'' each time series in time-direction (by attaching its end to the beginning)
randomly, thus destroying the inter correlation between the time series while
preserving the autocorrelation;
(2) calculating the CCC and its eigenvalues several times ($10^4 \sim 10^5$ times typically);
(3) comparing each distribution of the eigenvalues and the actual eigenvalue from the
largest in descending order: identifying the eigenmodes whose eigenvalue is larger
than that of the one obtained in the step (3) as significant modes.
This methodology overcomes the shortcoming of the RRS by allowing us to 
deal with data with nontrivial auto correlation and $T$ and $N$ not so large.

In this sense, the methodology of CHPCA with RRS is ideal for our purpose, which is to identify the customer choice process in our data.

\subsection{Results}

\begin{figure}
    \centering
    \includegraphics[width=0.6\textwidth]{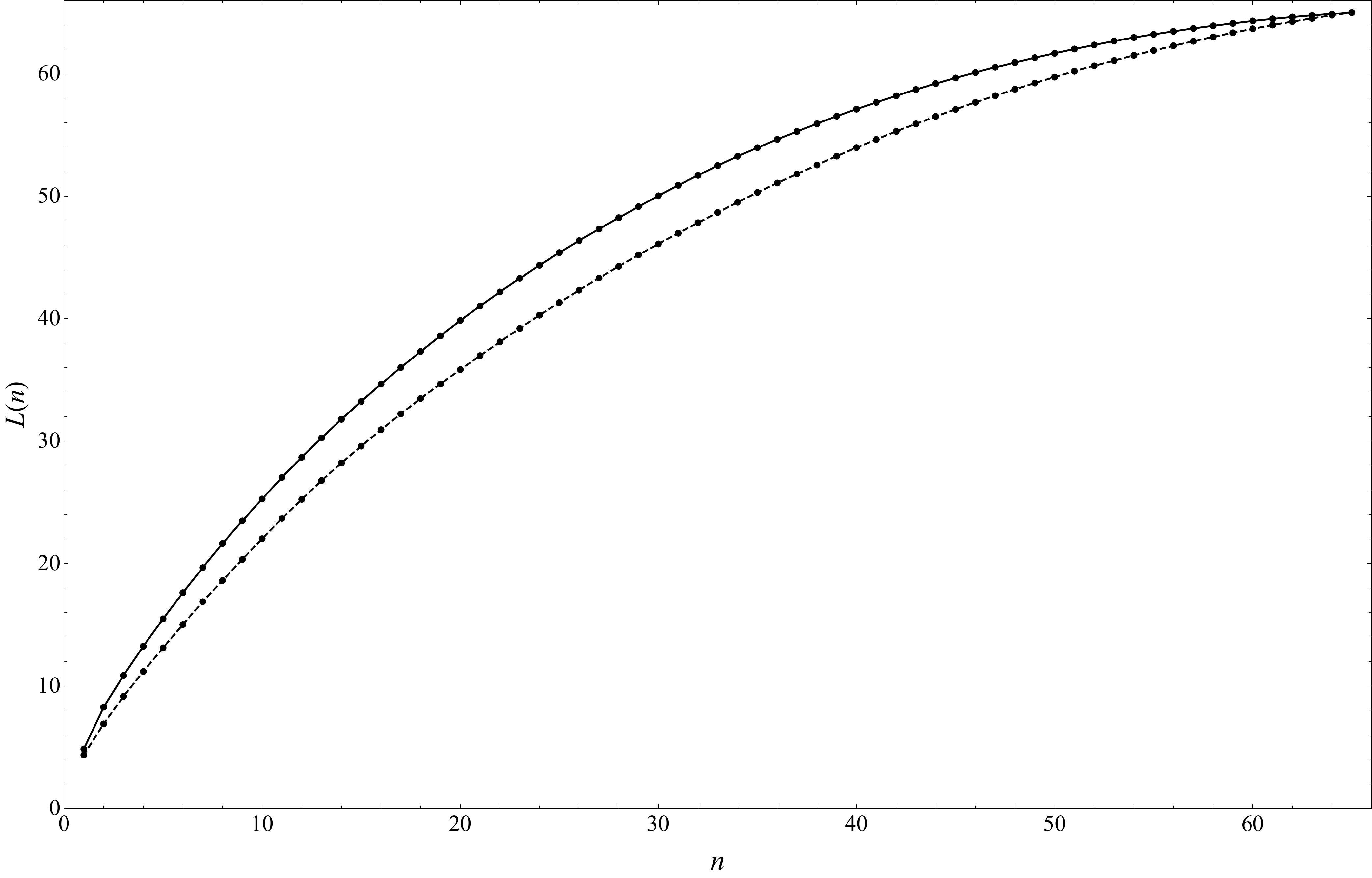}
    \caption{\bf Cumulative eigenvalue $L(n)$ defined in Eq.\eqref{eq:Ln}.}
    \label{fig:Beer-0}
\end{figure}

The eigenvalue distribution is shown in \figref{fig:Beer-0},
where the ordinate is the cumulative eigenvalue
\begin{equation}
    L(n):=\sum_{k=1}^n \lambda_n.
    \label{eq:Ln}
\end{equation}
The green dots are for CHPCA and blue for PCA.
As explained, the eigenvalue shows the rate of the presence of the corresponding eigenmode in the data. 
Therefore, this plot shows that CHPCA identifies the eigenmode more easily than PCA. This is natural since PCA misses movements with lead/lag.

The result of the RRS analysis of $10^4$ times the RRS simulation is summarized in \figref{fig:BeerRRS1} for the eigenvalues $n=1,2,3$ from top to bottom.
In each plot, the actual eigenvalue is shown by the thick vertical ticks.
The distribution shown in gray is the
distribution of the corresponding RRS eigenvalues, whose mean is shown by the
short vertical line, and the $2\sigma$ range is shown by the horizontal error bars. 
Since the eigenvalues \#1 and 2 are well above the $2\sigma$ range and \#3 is not, 
we find that the top two eigenmodes are significant, inter-correlating comovements.

\newcommand{\mw}{0.9\textwidth}
\begin{figure}
    \centering
    \includegraphics[width=\mw]{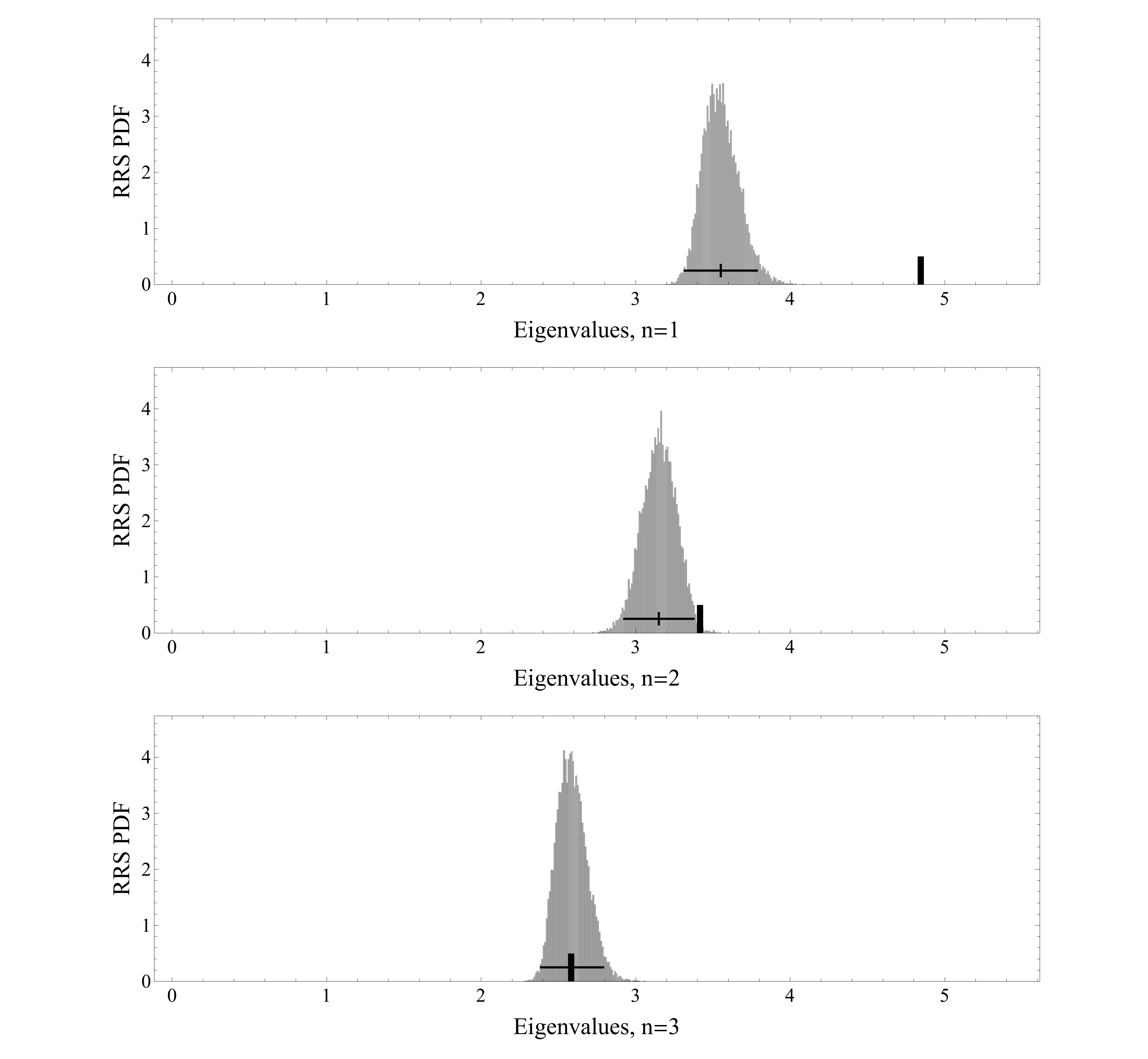}
    \caption{{\bf The eigenvalues (thick ticks) with the corresponding RRS eigenvalue distributions (shaded bell-shapes)
    and their 2$\sigma$ ranges (horizontal bars) for eigenvalues $n=1,2,3$.}
    The eigenvalues \#1 and \#2 are above the RRS 2$\sigma$ range and, therefore, are
    significant. The eigenvalues \#3 and below are not.}
    \label{fig:BeerRRS1}
\end{figure}

Since the largest two eigenvalues are
\begin{equation}
    \lambda_1=4.845 , \qquad
    \lambda_2=3.416 ,
    \label{eq:lambda12}
\end{equation}
respectively, 
these top eigenmodes take the share of 
$\sqrt{(\lambda_1+\lambda_2)/65}\simeq 0.3565$; {\it that is,}, 35.7\% of the
data are due to comovements.

\begin{figure}
    \centering
    \includegraphics[width=0.7\textwidth]{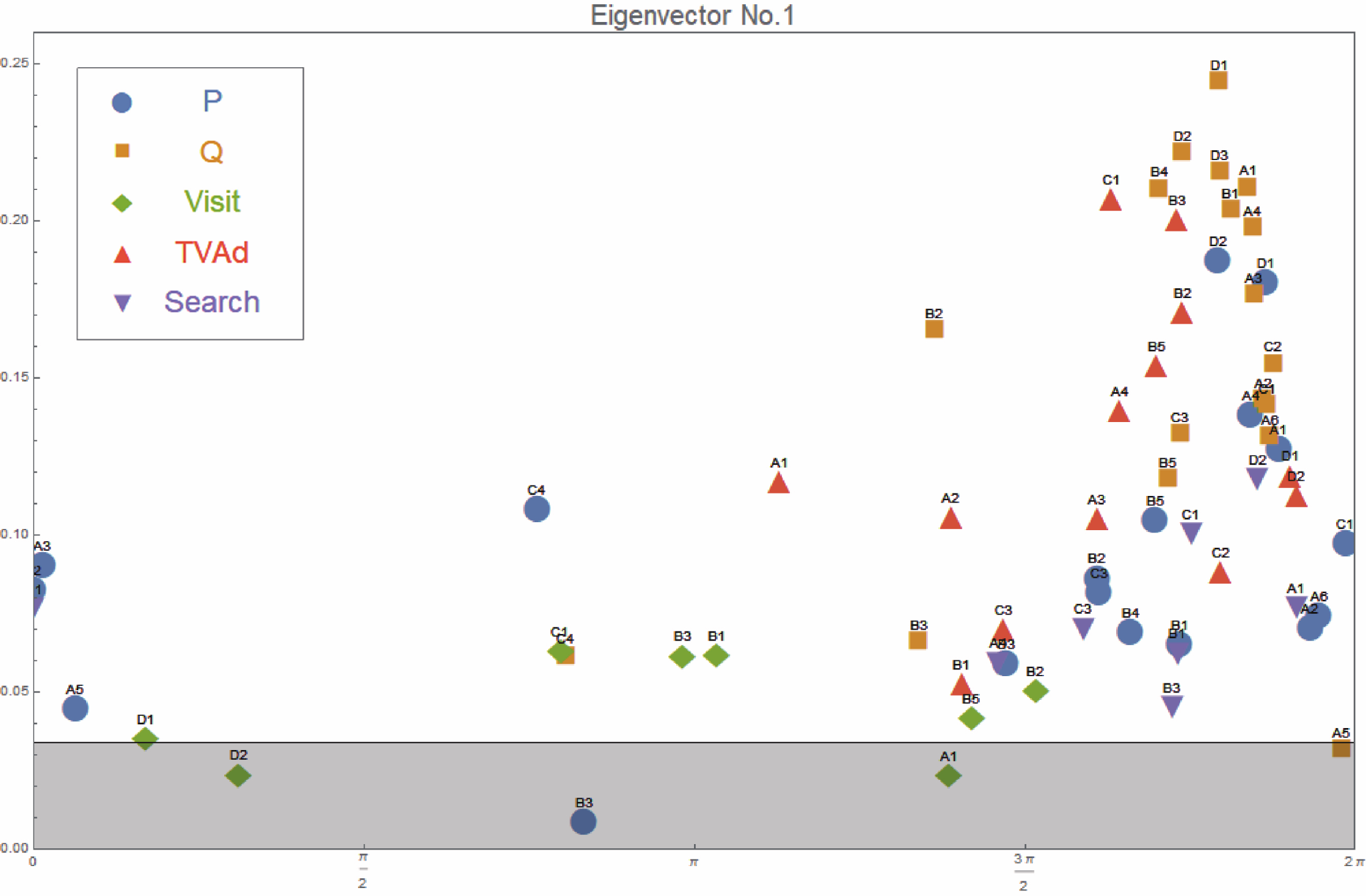}\\
    \includegraphics[width=0.7\textwidth]{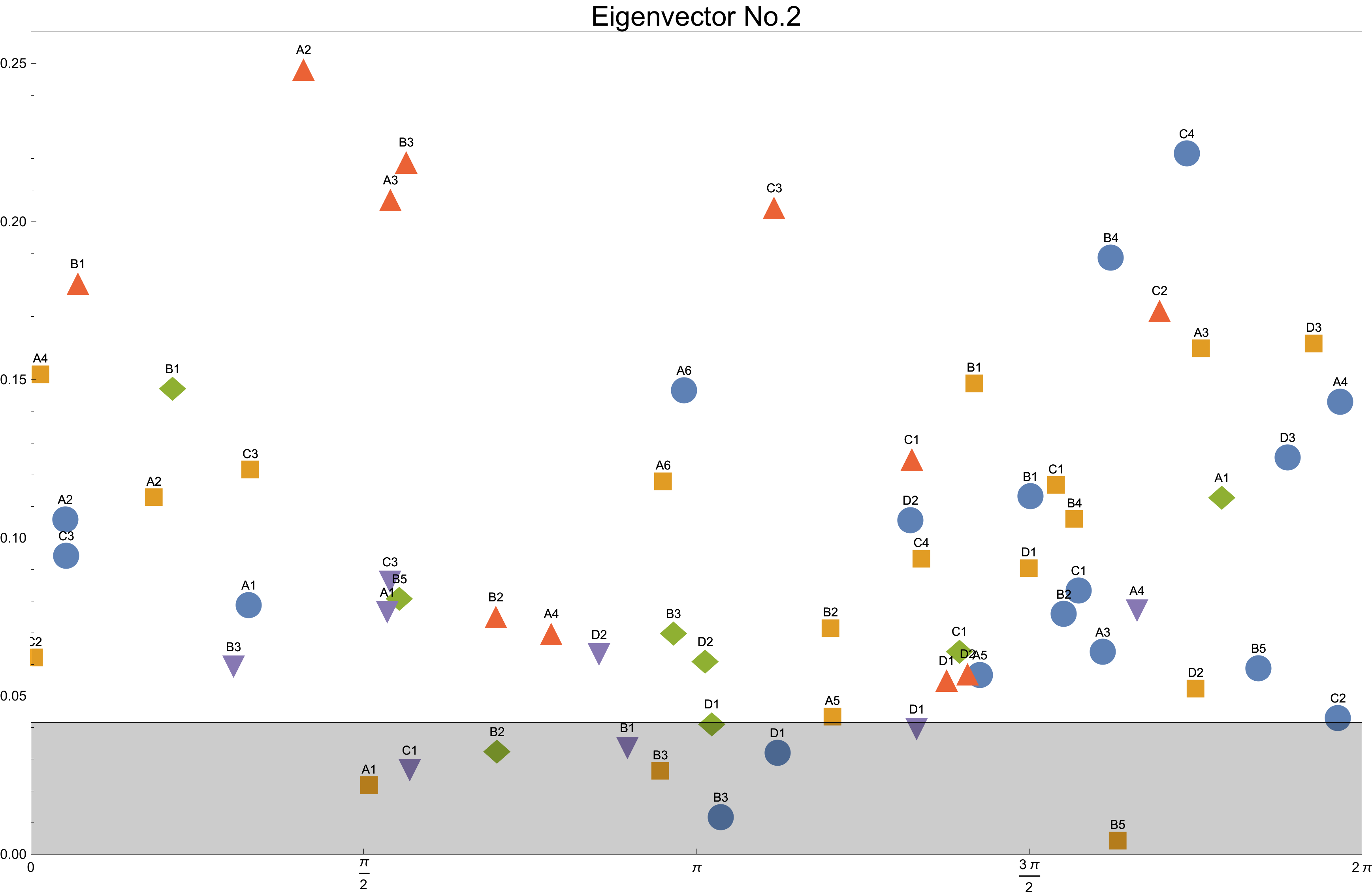}
    \caption{The components of the first (upper) and the second (lower) eigenvectors.}
    \label{fig:Beer-e}
\end{figure}

The top and the second eigenvector components are shown in \figref{fig:Beer-e}, where each component is shown by a marker specified by the product code at its top and the style shown in the legend.
(The details of these eigenmodes are given in the Appendix.)
The horizontal axis is its phase, and the vertical axis its absolute value.
The arbitrary overall phase in the eigenvector $\mathbf{e}_n$ is chosen so that the components representing purchase quantities are toward the right-hand side of the plots. 
By the definition of the complexification, the phase corresponds to the time-variation;
the components on the left move first, and the components to the right follow.
We have changed the phase of prices by $\pi$ to be consistent with the common knowledge that when the price goes down, quantity goes up.
In these plots, we also show the significance level of the absolute values of the components by the gray bands. Components with less absolute values have less significance in the respective eigenmode. To clarify this significance level, we add a random time series to the original data set and measure its absolute value in the first and second eigenvectors. Repeating this simulation 100 times, we identify the distribution of the absolute value. The shaded gray area bounded by a solid horizontal line is 2$\sigma$ range. Therefore, the components above the gray zones are the components with significant presence in the respective comovements.

The comovement of marketing instruments and purchase quantities across products represented in \figref{fig:Beer-e} may still be too complicated for marketers to interpret. Thus, we offer an additional method to reduce information obtained from CHPCA focusing on synchronization of multiple time series.

\clearpage

\section{Hodge Decomposition and Synchronization Network}
\subsection{Method}

The complex correlation coefficient, $C_{\alpha\beta}$, represents how strongly a pair of $\alpha$ and $\beta$ are correlated possibly with lead and lag. The strength of the correlation is given by the magnitude
\begin{equation}
  \label{eq:rho}
  \rho_{\alpha\beta}:=|C_{\alpha\beta}|\ ,
\end{equation}
and the lead and lag can be measured by the phase
\begin{equation}
  \label{eq:theta}
  \theta_{\alpha\beta}:=\arg C_{\alpha\beta}\ .
\end{equation}
Note that $\alpha$ leads $\beta$ if $\theta_{\alpha\beta}<0$,
and $\alpha$ lags $\beta$ if $\theta_{\alpha\beta}>0$ because we defined the direction of time by $e^{-i\omega t}$ (see Eq.(\ref{eq:iot})).

If we consider all the pairs in the complex correlation
$C_{\alpha\beta}$, we have a complete graph in which
every node $\alpha$ is connected to all the other nodes.
It is difficult to understand how individual
$\alpha$ leads or lags others in a more systematic way.
To overcome this difficulty, we select
\textit{comoving} pairs with strong correlation in the following way, and then use the so-called \textit{Hodge decomposition} of a flow on a directed and weighted network, which we call \textit{synchronization network}.

First, we select pairs of $\alpha$ and $\beta$ with
\begin{itemize}
    \item comovement: $0<\theta_{\alpha\beta}<\pi/2$,
    \item significant correlation: $\rho_{\alpha\beta}>\rho_*$
    where $\rho_*$ is a threshold given below.
\end{itemize}
In the first condition, we consider only the region $0<\theta_{\alpha\beta}<\pi/2$,
because the correlation matrix satisfies the Hermite conjugate relation; that is, 
$C_{\beta\alpha}=C_{\alpha\beta}^*$, so that the pairs in the region $-\pi/2<\theta_{\alpha\beta}<0$
are always in the region $0<\theta_{\alpha\beta}<\pi/2$.
In the second condition, we determine the threshold $\rho_*$ as follows.
If $\rho_*$ is too large, the number of pairs satisfying the condition is too small and, eventually, the graph becomes disconnected; if $\rho_*$ is too small, the graph is almost fully connected.
In both cases, it would be difficult to understand the lead/lag relation.
Therefore, we select $\rho_*$ that connects the graph at its largest value.
The resulting graph includes 65 nodes and 1,391 edges.

Second, we use a mathematical method of ranking nodes according to their location in terms of upstream and downstream flow in a directed network to
identify which nodes are leading and lagging in the entire relation.
In our case, a \textit{flow} is said to be present from $\alpha$ to $\beta$ if $0<\theta_{\beta\alpha}<\pi/2$
and $\rho_{\beta\alpha}=\rho_{\alpha\beta}>\rho_*$ with the amount of flow or weight, $\rho_{\alpha\beta}$.

We briefly recapitulate the method (see \cite{jiang2011hodge} for example),
which is called \textit{Hodge decomposition}.
Denote the adjacency matrix of the binary and weighted network by
\begin{equation}
  \label{eq:adjA}
  A_{\alpha\beta}=
  \begin{cases}
  1 & \text{if there is a directed edge from $\alpha$ to $\beta$},\\
  0 & \text{otherwise},
  \end{cases}
\end{equation}
and
\begin{equation}
  \label{eq:adjB}
  B_{\alpha\beta}=
  \begin{cases}
  f_{\alpha\beta} & \text{if there is a directed edge with a flow},\\
  0 & \text{otherwise},
  \end{cases}
\end{equation}
where $f_{\alpha\beta}$ is a flow from $\alpha$ to $\beta$,
and it is assumed that $f_{\alpha\beta}>0$.
Note that there can be such a pair of nodes that has
both $A_{\alpha\beta}=1$ and $A_{\beta\alpha}=1$ and also that has
both $f_{\alpha\beta}>0$ and $f_{\beta\alpha}>0$.

Then, the net \textit{flow} from $\alpha$ to $\beta$ is defined by
\begin{equation}
  \label{eq:matF}
  F_{\alpha\beta}=B_{\alpha\beta}-B_{\beta\alpha}\ .
\end{equation}
Let us also define the net \textit{weight} between $\alpha$ and $\beta$ by
\begin{equation}
  \label{eq:matW}
  W_{\alpha\beta}=A_{\alpha\beta}+A_{\beta\alpha}\ .
\end{equation}
Note that $F_{\alpha\beta}$ is anti-symmetric while
$W_{\alpha\beta}$ is symmetric.

Hodge decomposition is given by
\begin{equation}
  \label{eq:hodge_decomp}
  F_{\alpha\beta}=
  W_{\alpha\beta}\,(\phi_\alpha-\phi_\beta)
  +F_{\alpha\beta}^{\text{(loop)}}\ ,
\end{equation}
where $F_{\alpha\beta}^{\text{(loop)}}$ is a loop flow; that is,
divergence-free:
\begin{equation}
  \label{eq:hodge_Floop}
  \sum_\beta F_{\alpha\beta}^{\text{(loop)}}=0
\end{equation}
by definition. $\phi_\alpha$ is called \textit{Hodge potential}
of node $\alpha$.

Rewriting Eq.~\eqref{eq:hodge_Floop}, we have for each $\alpha=1,\cdots,N$,
\begin{equation}
  \label{eq:hodge_phi}
  \sum_\beta L_{\alpha\beta}\,\phi_\beta
  =\sum_\beta F_{\alpha\beta}\ ,
\end{equation}
Here, $L_{\alpha\beta}$ is the so-called graph Laplacian defined by
\begin{equation}
  \label{eq:hodge_L}
  L_{\alpha\beta}=\delta_{\alpha\beta}\sum_\gamma W_{\alpha\gamma}
  -W_{\alpha\beta}\ ,
\end{equation}
where $\delta_{\alpha\beta}=1$ if $\alpha=\beta$ and $\delta_{\alpha\beta}=0$ otherwise.
Given $F_{\alpha\beta}$ and $W_{\alpha\beta}$, Eqs.~\eqref{eq:hodge_phi}
are simultaneous linear equations to determine the Hodge potential $\phi_\alpha$ of all the nodes $\alpha$.

Note that simultaneous linear equations \eqref{eq:hodge_phi} are not independent of each other.
In fact, the summation over $\alpha$ of \eqref{eq:hodge_phi} is zero, as is easily shown,
corresponding to the fact that there is a freedom to fix
the origin of potential. It is not difficult to prove that
if the network is weakly connected; that is, connected
when considered an undirected graph, the potential can be
determined uniquely up to the choice of the origin \cite{USmacro}.
In the following, we use the convention that the mean is zero:
\begin{equation}
  \label{eq:hodge_phi_origin}
  \sum_\alpha \phi_\alpha=0\ .
\end{equation}

Thus, if we delete the loop flow, the remaining flow can
be represented by a flow caused by the difference in potential between
a pair of nodes. The Hodge potentials, therefore, can reveal which nodes are located in upstream or downstream sides in the relative relationship of the directed network. We emphasize that such information cannot be obtained
simply by looking at the pairwise correlation among nodes because the entire
connectivity of all the links is required to discard the loop flow
and to determine the potentials.

\subsection{Results and Interpretation}

\begin{figure}
  \centering
  \includegraphics[width=0.80\textwidth]{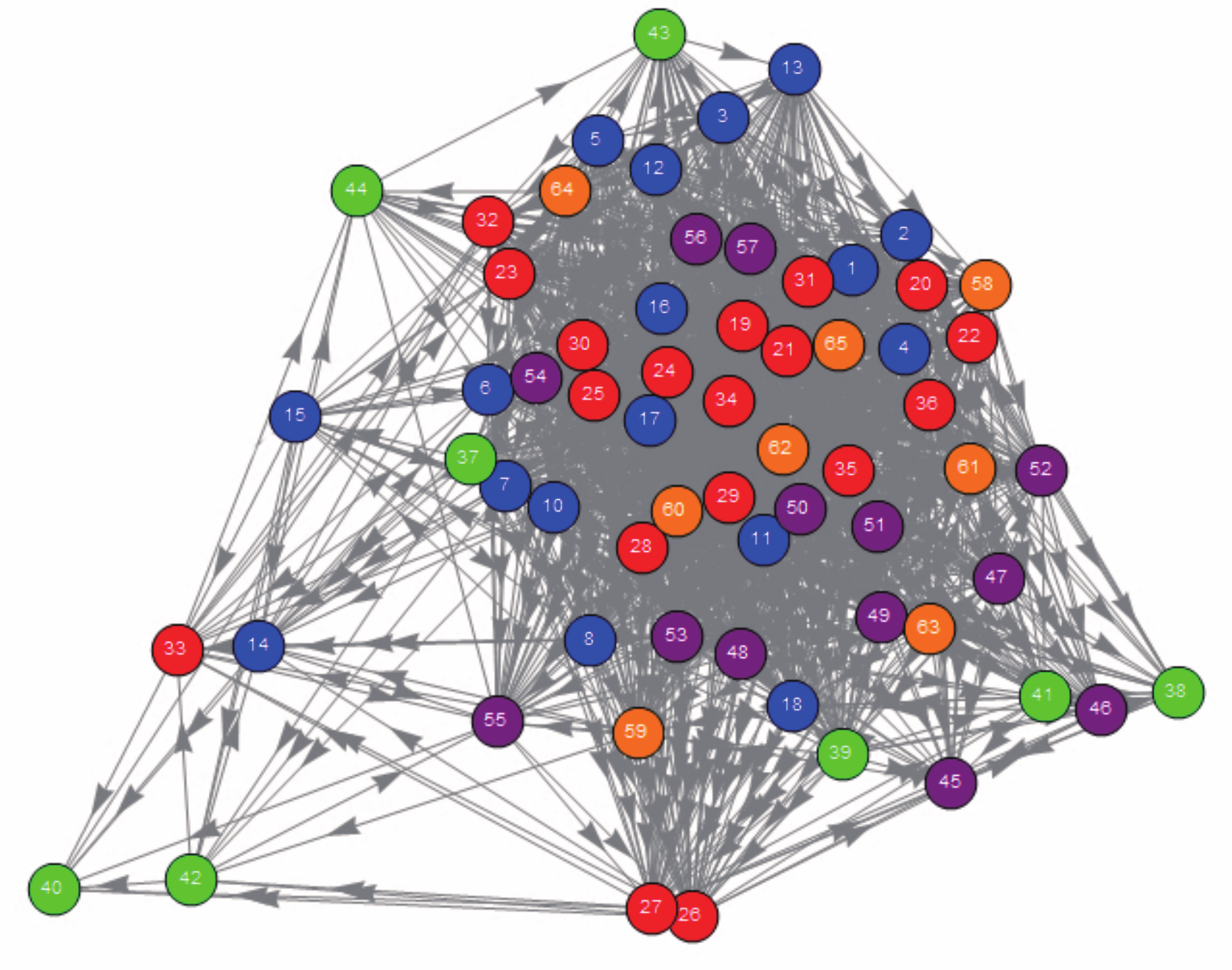}
  \caption{The synchronization network's graph layout.
    The vertical position of each node corresponds to its
    Hodge potential as a constraint in the force-directed
    algorithm of the graph layout. Upstream (leading) nodes
    are located toward the top while downstream (lagging)
    nodes are toward the bottom. The node no.9 is not drawn as it
    has only one link and is not relevant to this visualization.}
  \label{fig:SHN}
\end{figure}

\figref{fig:SHN} shows a layout of the synchronization network.
The vertical position of each node corresponds to its
Hodge potential as a constraint in the force-directed
algorithm of the graph layout. Upstream (leading) nodes
are located toward the top while downstream (lagging)
nodes are toward the bottom.

\begin{figure}
  \centering
  \includegraphics[width=0.95\textwidth]{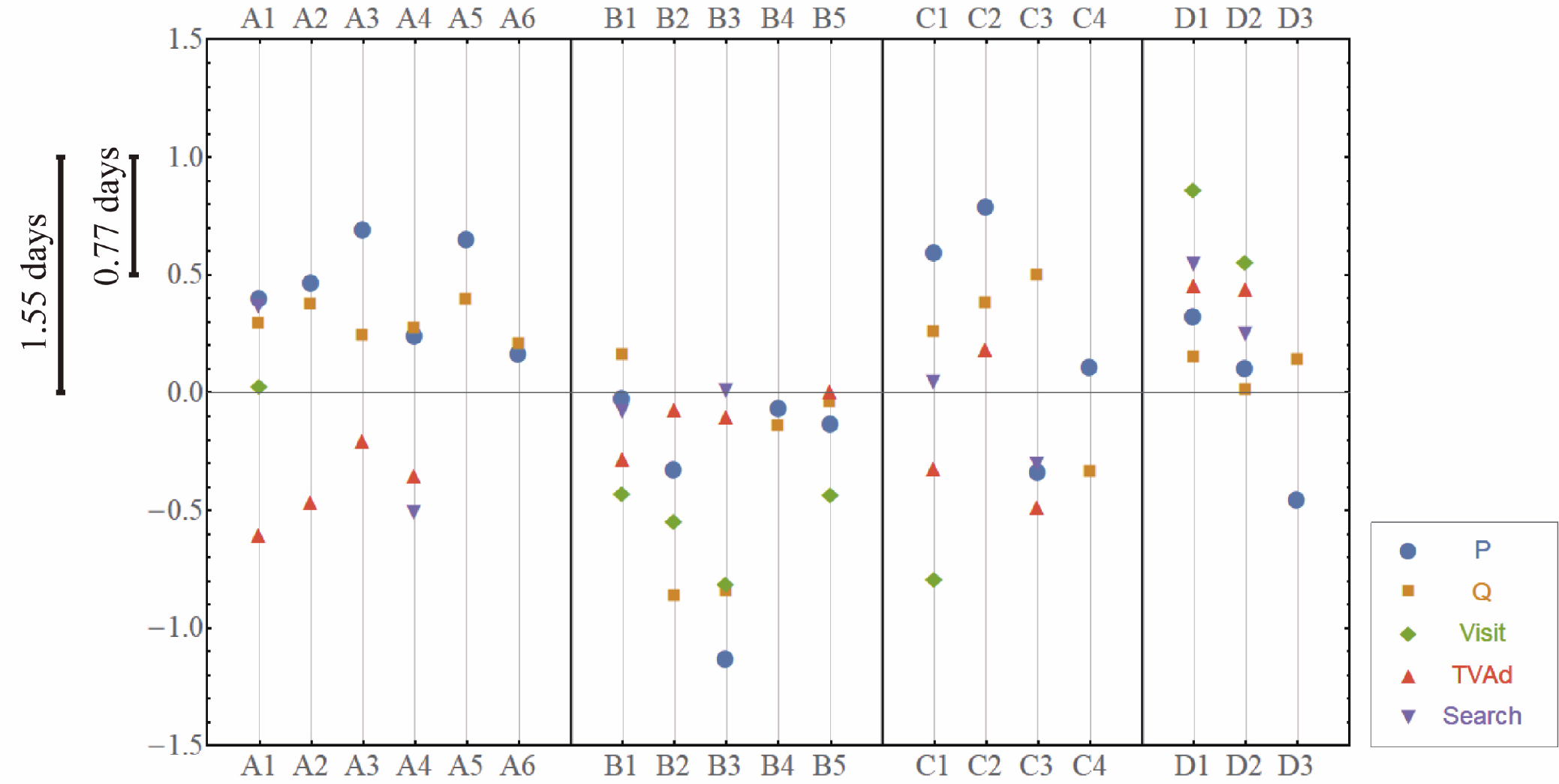}
  \caption{{\bf The fundamental sequence of exposures to marketing instruments and purchases for each product.} The vertical axis represents the Hodge potential, whose value is larger as it leads than others. Horizontally, products are arrayed in an arbitrary order. If the positions of two time series are closer, they are comoving almost simultaneously with each other. 
  }
  \label{fig:m1}
\end{figure}

To depict a customer choice process covering all competitive products, Hodge potentials are used to constitute the fundamental time sequence of the exposures to marketing instruments (TV Ad, web/mobile site visit, search and price) and the purchase quantity for each product. In Fig.\ref{fig:m1}, the time is passing from top to bottom along the vertical axis. The distance of this axis can be expressed in a time scale such that the distance of one corresponds to 1.55 days. The horizontal axis is nominal, where 18 products are arrayed in an arbitrary order. For instance, for Product 1, just after its price decreases and the search behavior increases, both of which occur almost simultaneously, the purchase quantity (sales) increases followed by an increase in exposure to web/mobile sites and TV advertising. 

First, we trace the time sequence of variables within each product. For more than half of the products (11 out of 18), a decrease (increase) in price leads to an increase (decrease) in quantity to a certain extent. For some products, a change in price occurs almost simultaneously with a change in quantity. For these two groups, the behaviors pf prices and quantities are consistent with standard economic theory. On the other hand, for Products 15 and 18, an increase (decrease) in quantity leads to a decrease (increase) in price. This phenomenon seems to be an anomaly as an effect of prices while it could be explained as an outcome of rational behavior; for instance, it could emerge when the demand is expanded by attracting new customers with lower willingness-to-pay \cite{kwon2018}.

The increased exposures to TV advertising lag behind the increased purchases for seven of the 13 products that executed TV advertising in the observed period. This may suggest that TV advertising by firms is a reaction to an increase in demand, not as an upfront investment or, in the latter case, after a significant time lag. The time sequence of web/mobile site visits or searches is less consistent across products than price or TV advertising. A possible reason is that the exposures to these marketing instruments is less controllable for firms, reflecting the idiosyncratic nature of individual products. In that sense, this inconsistency shows the advantage of our approach that analyzes the observed data purely empirically without any strong assumptions.

Second, we can compare the Hodge potentials horizontally between products, which indicates synchronization (comoving almost simultaneously) of marketing instruments between different products of a firm or even between firms. For instance, as Fig.\ref{fig:m1} shows, a price cut and sales increase for Products 1, 2 and 3, which are the main products of Firm 1, tend to be synchronized. That is, Firm 1 might coordinate price promotion consistently among their own products compared to rivals. These variables seem to be synchronized also between Firms 1 and 3, suggesting that these firms are mutually competing more intensively. As the potentials show that these firms tend to change prices before sales, their main weapon for competitive reaction is price promotion. 

It is noteworthy that the potentials for purchase quantity are relatively concentrated within the narrow band for most products, implying that beer consumption is highly synchronized as a whole. The reason is easily explained by the established knowledge that typical beer consumption increases during higher temperatures or special occasions such as weekends or holidays. A more interesting finding is the existence of a few products (Products 8 and 9 of Firm 2) outside the band. These products are interpreted as satisfying some niche demand in the market. Firm 2 seems to be differentiated since its pricing behavior is not necessarily synchronized with Firms 1 and 3 as a whole. Another prominent feature of this firm is that customers visited its web/mobile site more frequently while they seldom visited the sites of Firms 1 and 3 (or there may not be a competitor). Customers may visit Firm 2's site after making a purchase or exposure to TV advertising. On the other hand, customers seem to visit the site in advance. Such differences might be due to variations in marketing strategies.

\section{Customer Profile}\label{sec:application}

We found in 
the Method Section
that there are two significant eigenmodes
in the aggregate behavior of customers; the remaining $N-2$ modes can be discarded as ``noise.''
We are interested to see how the \textit{individual} behavior of customers
can be represented in terms of these two significant eigenmodes.
Such representation can provide deeper insight
into how the two significant eigenmodes can be interpreted by examining
individual customer's profiles such as their gender, age, income, 
other attributes, and their preferences for specific products.

Let us denote individual customer's time series by
$x_{p,\alpha}(t)$ ($p=1,\ldots,P$) where the index $p$ denotes
individual customers, and $P$ is the total number of customers in our data;
$P=1,738$.
$\alpha=1,\ldots,N$ is the same index as used in
the Method Section
with $N=65$.

We first complexify $x_{p,\alpha}(t)$ into complex
time series, denoted by $z_{p,\alpha}(t)$, and
standardize (subtract mean and normalize by standard deviation)
it precisely in the same way as we did in the Method  Section. 
Thus, we have
\begin{equation}
  \label{eq:def_zpa}
  \hat{z}_{p,\alpha}(t)=
  \frac{z_{p,\alpha}(t)-\langle z_{p,\alpha}\rangle}{\sigma_{p,\alpha}}.
\end{equation}
If $x_{p,\alpha}(t)$ is identically equal to zero
during all of time $t$ for some $\alpha$
(e.g., a customer was not exposed to any advertising),
we use the convention that $\hat{z}_{p,\alpha}(t)=0$.

Then, we project the time series to a space spanned by two significant eigenvectors (we term it \textit{customer space} hereafter); that is, \begin{equation}
  \label{eq:proj_a}
  a_{n,p}(t)=\sum_{\alpha=1}^N \,(\mathbf{e}_n)_\alpha^*\,\hat{z}_{p,\alpha}(t),
\end{equation}
for the two significant modes $n=1,2$ where $(\mathbf{e}_n)_\alpha$ is
the $\alpha$-th component of the eigenmode $\mathbf{e}_n$.
To locate each customer in a space spanned by the two eigenmodes,
we calculate the temporal mean of the squared magnitude of 
the projected time series $a_{n,p}(t)$, namely,
\begin{equation}
  \label{eq:proj}
  X_{n,p}=\frac{1}{T}\sum_t\left| a_{n,p}(t)\right|^2,
\end{equation}
which gives us two-dimensional coordinates for each customer $p$.

\begin{figure}
  \centering
  \includegraphics[width=0.6\textwidth]{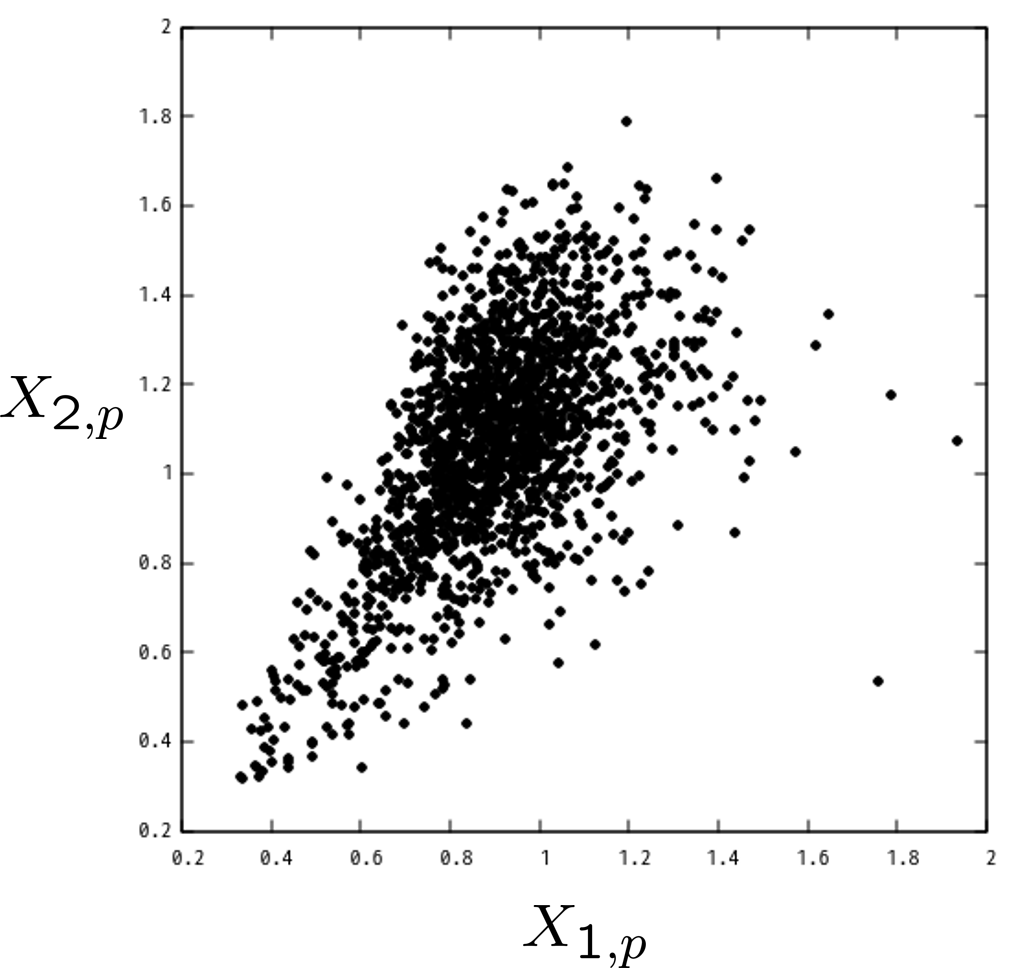}
  \caption{\bf Individual customer's projected representation
    for the two significant modes. See Eq.~\eqref{eq:proj}
    for the representation.}
  \label{fig:proj}
\end{figure}

\figref{fig:proj} shows the resulting spatial representation of Eq.~\eqref{eq:proj}
for all the $P$ customers.
Recalling \eqref{eq:eigenstrength} in
the Method Section,
each coordinate's value $X_{n,p}$ can be compared with
the eigenvalues $\lambda_1$ and $\lambda_2$, which are numerically given by Eq.~\eqref{eq:lambda12}.
We observe that there are customers whose positions are
consistent with Eq.~\eqref{eq:lambda12} in the sense that
$X_{1,p}/X_{2,p}\sim\lambda_1/\lambda_2$. There are, however,
more diversified customers in the two-dimensional space.
Such diversification tells us that the location of each customer
might be related to the heterogeneity in customer behavior.

\begin{table}
  \centering
    \caption{{\bf Regression of coordinates of customer space to customer profiles} (aa: $p<.001$; a: $p<.01$; b: $p<.05$; c: $p<.10$)}
    \scalebox{0.7}{
  \begin{tabular}{l|rrl|rrl|rrl|rrl}
    \hline
    & \multicolumn{3}{c|}{Model 1.1} & \multicolumn{3}{c|}{Model 1.2} & \multicolumn{3}{c|}{Model 2.1} & \multicolumn{3}{c}{Model 2.2} \\
    \hline
    Criterion Variable: & \multicolumn{6}{c|}{coord. for 1st eigen mode: $X_{1,p}$} & \multicolumn{6}{c}{coord. for 2nd eigen mode: $X_{2,p}$} \\
    \hline
    Explanatory Variables: & coef. & s.e. & & coef. & s.e. & & coef. & s.e. & & coef. & s.e. & \\
    \hspace{1em}Intercept & $.9050$ & $.0045$ & {\footnotesize aa} & $.9050$ & $.0044$ & {\footnotesize aa} & $1.0720$ & $.0058$ & {\footnotesize aa} & $1.0720$ & $.0058$ & {\footnotesize aa} \\
    \hspace{1em}Age & $.0164$ & $.0048$ & {\footnotesize aa} & $.0159$ & $.0047$ & {\footnotesize aa} & $.0347$ & $.0061$ & {\footnotesize aa}& $.0352$ & $.0061$ & {\footnotesize aa} \\
    \hspace{1em}Gender & $.0033$ & $.0059$ & & $.0040$ & $.0059$ & & $.0036$ & $.0077$ & & $.0036$ & $.0077$ & \\
    \hspace{1em}Marrital Status & $.0013$ & $.0050$ & & $.0007$ & $.0049$ & & $.0030$ & $.0064$ & & $.0018$ & $.0064$ & \\
    \hspace{1em}Personal Income & $.0071$ & $.0065$ & & $.0084$ & $.0065$ & & $-.0099$ & $.0084$ & & $-.0092$ & $.0084$ & \\
    \hspace{1em}Houshold Income & $-.0011$ & $.0055$ & & $-.0034$ & $.0055$ & & $-.0067$ & $.0071$ & & $-.0078$ & $.0072$ & \\
    \hspace{1em}Total Purchase (freq.) & $.0351$ & $.0055$ & {\footnotesize aa} & $.0331$ & $.0056$ & {\footnotesize aa} & $.0257$ & $.0070$ & {\footnotesize aa} & $.0273$ & $.0073$ & {\footnotesize aa} \\
    \hspace{1em}Total Purchase (m$\ell$) & $.0139$ & $.0055$ & {\footnotesize b} & --- & --- & & $.0165$ & $.0071$ & {\footnotesize b} & --- & --- & \\
    \hspace{1em}Purchase -- A1 & & & & $.0091$ & $.0047$ & {\footnotesize c} & & & & $-.0007$ & $.0060$ & {\footnotesize c} \\
    \hspace{1em}\phantom{Purchase -- }A2 & & & & $.0028$ & $.0045$ & & & & & $.0082$ & $.0058$ & \\
    \hspace{1em}\phantom{Purchase -- }A3 & & & & $.0035$ & $.0045$ & & & & & $.0066$ & $.0058$ & \\
    \hspace{1em}\phantom{Purchase -- }A4 & & & & $.0031$ & $.0047$ & & & & & $.0086$ & $.0061$ & \\
    \hspace{1em}\phantom{Purchase -- }A5 & & & & $.0059$ & $.0045$ & & & & & $.0107$ & $.0058$ & \\
    \hspace{1em}\phantom{Purchase -- }A6 & & & & $-.0005$ & $.0047$ & & & & & $.0007$ & $.0060$ & \\
    \hspace{1em}\phantom{Purchase -- }B1 & & & & $.0044$ & $.0046$ & & & & & $.0008$ & $.0059$ & \\
    \hspace{1em}\phantom{Purchase -- }B2 & & & & $.0046$ & $.0045$ & & & & & $-.0013$ & $.0058$ & \\
    \hspace{1em}\phantom{Purchase -- }B3 & & & & $-.0058$ & $.0045$ & & & & & $-.0048$ & $.0059$ & \\
    \hspace{1em}\phantom{Purchase -- }B4 & & & & $-.0002$ & $.0045$ & & & & & $.0070$ & $.0058$ & \\
    \hspace{1em}\phantom{Purchase -- }B5 & & & & $.0117$ & $.0047$ & {\footnotesize b} & & & & $.0092$ & $.0061$ & \\
    \hspace{1em}\phantom{Purchase -- }C1 & & & & $-.0073$ & $.0046$ & & & & & $-.0033$ & $.0060$ & \\
    \hspace{1em}\phantom{Purchase -- }C2 & & & & $-.0006$ & $.0045$ & & & & & $-.0037$ & $.0059$ & \\
    \hspace{1em}\phantom{Purchase -- }C3 & & & & $.0035$ & $.0046$ & & & & & $.0162$ & $.0059$ & {\footnotesize aa} \\
    \hspace{1em}\phantom{Purchase -- }C4 & & & & $.0067$ & $.0045$ & & & & & $.0073$ & $.0059$ & \\
    \hspace{1em}\phantom{Purchase -- }D1 & & & & $.0168$ & $.0046$ & {\footnotesize aa} & & & & $.0031$ & $.0059$ & \\
    \hspace{1em}\phantom{Purchase -- }D2 & & & & $-.0014$ & $.0048$ & & & & & $-.0048$ & $.0062$ & \\
    \hspace{1em}\phantom{Purchase -- }D3 & & & & $.0170$ & $.0045$ & {\footnotesize aa} & & & & $.0104$ & $.0058$ & \\
    \hline
    $\text{R}^2$ & \multicolumn{3}{c|}{$.0650$} & \multicolumn{3}{c|}{$.0870$} & \multicolumn{3}{c|}{$.0484$} & \multicolumn{3}{c}{$.0870$} \\
    $\text{Adjusted R}^2$ & \multicolumn{3}{c|}{$.0612$} & \multicolumn{3}{c|}{$.0742$} & \multicolumn{3}{c|}{$.0445$} & \multicolumn{3}{c}{$.0742$} \\
    \hline
  \end{tabular}}
  \label{tab:regprof}
\end{table}

To assess how the customer space is associated with each customer's profile, we conduct regression analysis where either of the coordinates in the customer space, $X_{1,p}$ or $X_{2,p}$, is used as a criterion variable, and multiple variables representing customer profiles are used as explanatory variables, which are all available in our data set. First, we select age (nine-point scale from age 20 to 24 to age 60 or older), gender (0 for male, 1 for female), marital status (0 for unmarried, 1 for the married), personal income (nine-point scale), and household income (five-point scale) by preliminary analysis. Second, to capture each customer's beer preference, total purchase frequency and quantity (ml) over all products are added to the predictors. As the results of Model 1.1 and 2.1 of Table 4 show, the estimated coefficients are significant only for age (0.1\% significant), total purchase frequency (0.1\% significant), and quantity (5\% significant) for both $X_{1,p}$ or $X_{2,p}$. The values of $R^2$ indicate that most of the variations are not explained by the above predictors.

Alternatively, we replace total purchase quantity with each product's purchase quantities to capture individual differences in product-level preference. The results are presented for Model 1.2 and 2.2 in Table 4. Compared to the above models, the coefficients for age and total purchase frequency are consistently significant while the adjusted $R^{2}$s are slightly increased, implying separating total purchase quantity into purchase quantity for products may marginally improve the model fit. The coefficients of a few products are significant from the 0.1\% to 10\% level for $X_{1,p}$ and $X_{2,p}$. Thus, locations in the customer space could be explained to some extent by age, purchase frequency at the category level, and purchases of some remarkable products. However, it should be noted that most of the variations in the customer space remain unexplained. In other words, the individual locations in a customer space might reflect an infinite number of factors, only some of which could be measured via customer surveys or purchase history tracking. Hence, our proposed projection method contributes to the evaluation of individual deviations from a representative customer choice process.

\section{Conclusion}\label{sec:conclusion}

The proliferation of marketing instruments and competitive products are rendering the consumer choice process increasingly complex to define. Existing methods used for this purpose ignore the existence of competitive products or underestimate the range of competition, even in cases where most customers consider multiple alternative products by searching or shopping. One of the main reasons for this is the limited ability of the existing methods, such as multivariate time series analyses, to handle the complexity caused by multiple products competing with multiple instruments. The market with 18 products and 5 variables (4 marketing instruments and one purchase quantity), analyzed in this study, is not tractable without strong assumptions to drastically reduce parameters. Our proposed CHPCA overcomes this limitation without any strong assumptions. Furthermore, CHPCA's supplemental methods, synchronization network and Hodge decomposition, can be used to summarize and visualize the results to be more interpretable.  

This study shows that a set of our proposed methods could be used to effectively understand the consumer choice process embedded in enormously high-dimensional time-series data. The application of our method to the beer market in Japan derives some interesting findings. First, for most products on the market, the increase (decrease) in a product's price leads to the decrease (increase) in its purchase quantity (as the standard economic theory predicts) with a few exceptional cases. Second, the exposure to TV advertising increases later than the purchase quantity in many cases. Simply put, consumers notice a price change in a store, buy a product, and are then exposed to TV ads later. Third, the timing of consumers' visits to product web/mobile sites or the usage of search engines varies across products. Fourth, we find synchronization across products, in particular within each firm, rather than across firms. These findings imply that individual firms are heterogeneous, each adopting a distinctive coherent marketing strategy.  

The fourth point has an important implication for economic policy making. Synchronization of marketing strategy between firms indicates that their behavior could be competitive if prices are decreasing but be collusive if prices are increasing. The latter case should attract a strong interest of anti-trust agencies. Our result might deny this possibility, while suggesting another difficulty in economic policy making. If corporate behaviors are heterogeneous than expected, policymakers must allow for such heterogeneity in evaluating the effectiveness of planned policies in advance. Firms should not be treated as an aggregation of representative agents.

For managers in firms, the above-mentioned findings may be instructive to improve their marketing practices. If TV advertising reaches customers later than their purchase, the timing of advertisement should be reconsidered, since consumers cannot choose the timing of exposure to it. On the other hand, if the ad campaign intends to reinforce customer retention, the marketing practice could be successful. Our method reveals which products could be real rivals, without any prejudice, by showing synchronization of marketing instruments between competitive products. This information can be used by marketers to investigate the dynamics of competition or substitution for their product.

Researches in marketing science have traditionally measured brand loyalty by incorporating own-product inertia \cite{guadagni1983} and the long-term effect of advertising by incorporating ad stock into the model \cite{nerlove1962}. Our method could implicitly capture the effect of brand loyalty that is embedded in the comovement of marketing instruments. How to explicitly quantify the magnitude of loyalty or inertia may be a possible challenge for us. We have already attempted incorporating ad stock with exponentially-distributed weights and alternative parameters. As the result was not sensitive to such modification, we tentatively conclude that accounting for the long-term effects of advertising in our method is not a priority.   

In our opinion, it is desirable to develop further extension of our method, aiming to offer some numerical indications for policymakers and marketers, to enable better actions, by proceeding to sensitivity analysis/simulations, using the synchronization structure discussed in this paper. 
A fluctuation-dissipation approach \cite{iyetomi2011fluctuation} may be useful, assuming that the impact of external stimuli does not change the correlation structure, but simply excites some of the structure. In other words, this approach deals with small perturbations on the existing structure, which is, in general, true when promoting specific products or regulating specific firm's behavior. Research in this direction, therefore, would be fruitful.


\section*{Acknowledgments}

This study is an outcome of the Project ``Dynamics of Economy and
Finance from the Economic Network Point of View'' at the Research
Institute of Economy, Trade and Industry (RIETI). %
This work is partially supported by Grant-in-Aid for Scientific
Research (KAKENHI) of JSPS Grant Numbers 17H02041, 20H02391
and %
by MEXT as Exploratory Challenges on Post-K Computer (Studies of
Multilevel Spatiotemporal Simulation of Socioeconomic Phenomena:
Macroeconomic Simulations). The authors appreciate the support by
INTAGE Inc.\ to make the data available. The authors would also like
to thank participants of the conferences of Computational Social
Science Japan, Japan Institute of Marketing Science, and Japan Society
for Evolutionary Economics for their helpful discussions and comments.
We would like to thank Editage (www.editage.com) for English language
editing.


\begin{thebibliography}{10}

\bibitem{edelman2015}
Edelman DC, Singer M.
\newblock Competing on customer journeys.
\newblock Harvard Business Review. 2015;93(11):88--100.

\bibitem{lemon2016}
Lemon KN, Verhoef PC.
\newblock Understanding Customer Experience Throughout the Customer Journey.
\newblock Journal of Marketing. 2016;80(11):69--96.

\bibitem{anderl2016}
Anderl E, Becker I, von Wangenheim F, Schumann JH.
\newblock Mapping the Customer Journey: Lessons Learned from Graph-based Online
  Attribution Modeling.
\newblock International Journal of Research in Marketing. 2016;33(3):457--474.

\bibitem{dehaan2015}
{De Haan} E, Kannan PK, Verhoef PC, Wiesel T. The Role of Mobile Devices in the
  Online Customer Journey; 2015.

\bibitem{pauwels2007}
Pauwels K.
\newblock How Retailer and Competitor Decisions Drive The Long-Term
  Effectiveness of Manufacturer Promotions for Fast Moving Consumer Goods.
\newblock Journal of Retailing. 2007;83(3):297--308.

\bibitem{ataman2010}
Ataman MB, Van~Heerde HJ, Mela CF.
\newblock The long-term effect of marketing strategy on brand sales.
\newblock Journal of Marketing Research. 2010;47(5):866--882.

\bibitem{sriram2007}
Sriram S, Balachander S, Kalwani MU.
\newblock Monitoring the Dynamics of Brand Equity Using Store-Level Data.
\newblock Journal of Marketing. 2007;71(4):61--78.

\bibitem{kolsarici2016}
Kolsarici C, Vakratsas D.
\newblock Correcting for Misspecification in Parameter Dynamics to Improve
  Forecast Accuracy with Adaptively Estimated Models.
\newblock Management Science. 2016;61(10):2495--2513.

\bibitem{steenkamp2005}
Steenkamp JBEM, Nijs VR, Hanssens DM, Dekimpe MG.
\newblock Competitive Reactions to Advertising and Promotion Attacks.
\newblock Marketing Science. 2005;24(1):35--54.

\bibitem{aoyama2010econophysics}
Aoyama H, Fujiwara Y, Ikeda Y, Iyetomi H, Souma W.
\newblock Econophysics and Companies: Statistical Life and Death in Complex
  Business Networks.
\newblock Cambridge University Press; 2010.

\bibitem{aoyamacambridge2017}
Aoyama H, Fujiwara Y, Ikeda Y, Iyetomi H, Yoshikawa H.
\newblock Macro-Econophysics, New Studies on Economic Networks and
  Synchronization.
\newblock Cambridge University Press; 2017.

\bibitem{kichikawa2020ilp}
Kichikawa Y, Iyetomi H, Aoyama H, Fujiwara Y, Yoshikawa H.
\newblock Interindustry Linkages of Prices: Analysis of Japan's Deflation.
\newblock PLoS ONE. 2020;15(2):e0228026.

\bibitem{vodenska2016}
Vodenska I, Aoyama H, Fujiwara Y, Iyetomi H, Arai Y.
\newblock Interdependencies and causalities in coupled financial networks.
\newblock PloS one. 2016;11(3):e0150994.

\bibitem{hartman2011}
Hartmann W, Nair HS, Narayanan S.
\newblock Identifying Causal Marketing Mix Effects Using a Regression
  Discontinuity Design.
\newblock Marketing Science. 2011;30(6):1079--1097.

\bibitem{mizuno2006}
Mizuno M, Hoshino T. Assessing the Short-term Causal Effect of TV Advertising
  via the Propensity Score Method; 2006.

\bibitem{RACJ1981}
Rasmusson EM, Arkin PA, Chen WY, Jalickee JB.
\newblock Biennial variations in surface temperature over the United States as
  revealed by singular decomposition.
\newblock Mon Wea Rev. 1981;109:587--598.

\bibitem{Barnett1983}
Barnett TP.
\newblock Interaction of the monsoon and Pacific trade wind system at
  interannual time scales part I: The equatorial zone.
\newblock Mon Wea Rev. 1983;111:756--773.

\bibitem{Horel1984}
Horel JD.
\newblock Complex principal component analysis: Theory and examples.
\newblock J Appl Meteor. 1984;23:1660--1673.

\bibitem{PhysRevLett.107.128501}
Stein K, Timmermann A, Schneider N.
\newblock Phase synchronization of the El Ni\~no-Southern oscillation with the
  annual cycle.
\newblock Phys Rev Lett. 2011;107:128501.
\newblock doi:{10.1103/PhysRevLett.107.128501}.

\bibitem{JOC:JOC1499}
Hannachi A, Jolliffe IT, Stephenson DB.
\newblock Empirical orthogonal functions and related techniques in atmospheric
  science: A review.
\newblock International Journal of Climatology. 2007;27(9):1119--1152.
\newblock doi:{10.1002/joc.1499}.

\bibitem{hilbert}
Hilbert D.
\newblock Grundz\"ge einer allgemeinen theorie der linearen
  integralgleichungen.
\newblock Druck und Verlag con B.\ G.\ Teubner; 1912.

\bibitem{Gabor1946}
Gabor D.
\newblock Theory of communication.
\newblock J Inst Electr Eng--Part III, Radio Commun Eng. 1946;93:429--457.

\bibitem{granger1964spectral}
Granger CWJ, Hatanaka M.
\newblock Spectral analysis of economic time series.
\newblock Princeton Univ. Press.; 1964.

\bibitem{bendat2011random}
Bendat JS, Piersol AG.
\newblock Random Data: Analysis and Measurement Procedures.
\newblock Wiley Series in Probability and Statistics. Wiley; 2011.
\newblock Available from:
  \url{http://books.google.co.jp/books?id=iu7pq6\_vo3QC}.

\bibitem{feldman2011hilbert}
Feldman M.
\newblock Hilbert transform in vibration analysis.
\newblock Mechanical systems and signal processing. 2011;25(3):735--802.

\bibitem{hilbertikeda}
Ikeda Y, Aoyama H, Yoshikawa H.
\newblock Synchronization and the coupled ocsillator model in international
  business cycles.
\newblock RIETI discussion paper. 2013;13-E-086.

\bibitem{ikeda2013direct}
Ikeda Y, Aoyama H, Iyetomi H, Yoshikawa H.
\newblock Direct evidence for synchronization in Japanese business cycles.
\newblock Evolutionary and Institutional Economics Review. 2013;10(2):315--327.

\bibitem{iyetomiarai2013}
Arai Y, Iyetomi H.
\newblock Complex principal component analysis of dynamic correlations in
  financial markets.
\newblock Intelligent Decision Technologies, Frontiers in Artificial
  Intelligence and Applications. 2013;255:111--119.

\bibitem{jiang2011hodge}
Jiang X, Lim LH, Yao Y, Ye Y.
\newblock Statistical Ranking and Combinatorial Hodge Theory.
\newblock Mathematical Programming. 2011;127(1):203--244.

\bibitem{USmacro}
Iyetomi H, Aoyama H, Fujiwara Y, Souma W, Vodenska I, Yoshikawa H.
\newblock Relationship between Macroeconomic indicators and economic cycles in
  {U.S.}
\newblock Scientific Reports. 2020;10(1):1--12.

\bibitem{kwon2018}
Kwon M, Erdem T, Ishihara M. Counter-Cyclical Price Promotion: Capturing
  Seasonal Category Expansion Under Endogenous Consumption; 2018.

\bibitem{guadagni1983}
Guadagni PM, Little JD.
\newblock A Logit Model of Brand Choice Calibrated on Scanner Data.
\newblock Marketing Science. 1983;2(3):203--238.

\bibitem{nerlove1962}
Nerlove M, Arrow KJ.
\newblock Optimal Advertising Policy under Dynamic Conditions.
\newblock Economica. 1962;29(114):129--142.

\bibitem{iyetomi2011fluctuation}
Iyetomi H, Nakayama Y, Aoyama H, Fujiwara Y, Ikeda Y, Souma W.
\newblock Fluctuation-dissipation theory of input-output interindustrial
  relations.
\newblock Physical Review E. 2011;83(1):016103.

\end{thebibliography}

\clearpage
\section*{APPENDIX: The eigenmodes 1 and 2: detail}

We list the components of the first eigenmode with absolute value above the 2$\sigma$ range in \figref{tab:A1left} and \figref{tab:A1right} and, similarly, for the second eigenmode in \figref{tab:A2left} and \figref{tab:A2right}.

\begin{table}[h]
  \begin{minipage}[t]{.45\textwidth}
    \begin{center}
     \scalebox{0.7}{
      \begin{tabular}{c|c|cc}
\hline
Brand&Variable&Phase&Abs.\\
\hline
 \text{A1} & \text{TVAd} & 2.22 & 0.21 \\
 \text{A1} & \text{WebV} & 2.57 & 0.07 \\
 \text{A1} & \text{MobV} & 2.85 & 0.10 \\
 \text{A1} & \text{Q} & 4.91 & 0.12 \\
 \text{A1} & \text{P} & 5.09 & 0.08 \\
 \hline
 \text{A2} & \text{WebV} & 2.90 & 0.14 \\
 \text{A2} & \text{TVAd} & 2.93 & 0.09 \\
 \text{A2} & \text{MobV} & 3.08 & 0.08 \\
 \text{A2} & \text{P} & 5.57 & 0.07 \\
 \hline
 \text{A3} & \text{MobV} & 2.59 & 0.18 \\
 \text{A3} & \text{WebV} & 2.8 & 0.20 \\
 \text{A3} & \text{TVAd} & 2.93 & 0.08 \\
 \text{A3} & \text{Q} & 4.83 & 0.09 \\
 \text{A3} & \text{P} & 5.32 & 0.08 \\
 \hline
 \text{A4} & \text{TVAd} & 0.44 & 0.08 \\
 \text{A4} & \text{Q} & 5.29 & 0.07 \\
 \hline
 \text{A6} & \text{P} & 3.88 & 0.10 \\
 \text{A6} & \text{Q} & 5.49 & 0.11 \\
 \hline
 \text{A7} & \text{P} & 4.85 & 0.11 \\
 \text{A7} & \text{Q} & 5.99 & 0.14 \\
 \hline
 \text{B1} & \text{WebV} & 1.82 & 0.07 \\
 \text{B1} & \text{TVAd} & 2.12 & 0.08 \\
 \text{B1} & \text{Q} & 5.41 & 0.09 \\
 \text{B1} & \text{P} & 5.83 & 0.07 \\
 \hline
 \text{B2} & \text{Q} & 3.90 & 0.08 \\
 \text{B2} & \text{TVAd} & 4.81 & 0.08 \\
 \hline
 \text{B3} & \text{WebS} & 3.27 & 0.07 \\
 \text{B3} & \text{TVAd} & 5.39 & 0.14 \\
 \hline
 \text{B4} & \text{Q} & 4.91 & 0.09 \\
 \hline
      \end{tabular}}
    \end{center}
    \caption{List of components in the first eigenmode with absolute value above the 2$\sigma$ range
    (to be continued to \figref{tab:A1right})}
    \label{tab:A1left}
  \end{minipage}
  %
  %
  \begin{minipage}[t]{.45\textwidth}
    \begin{center}
      \scalebox{0.7}{
      \begin{tabular}{c|c|cc}
\hline
Brabd&Variable&Phase&Abs.\\
\hline
 \text{B5} & \text{WebV} & 3.52 & 0.08 \\
 \text{B5} & \text{TVAd} & 4.18 & 0.15 \\
 \text{B5} & \text{P} & 4.21 & 0.08 \\
 \text{B5} & \text{Q} & 5.03 & 0.07 \\
 \hline
 \text{C1} & \text{WebV} & 3.62 & 0.07 \\
 \text{C1} & \text{TVAd} & 4.57 & 0.11 \\
 \text{C1} & \text{Q} & 5.33 & 0.08 \\
 \hline
 \text{C2} & \text{MobS} & 1.86 & 0.10 \\
 \text{C2} & \text{MobV} & 1.86 & 0.10 \\
 \text{C2} & \text{Q} & 5.63 & 0.11 \\
 \text{C2} & \text{P} & 6.26 & 0.10 \\
 \hline
 \text{C3} & \text{TVAd} & 5.50 & 0.12 \\
 \text{C3} & \text{Q} & 5.89 & 0.25 \\
 \text{C3} & \text{P} & 6.16 & 0.27 \\
 \hline
 \text{C4} & \text{Q} & 0.96 & 0.11 \\
 \text{C4} & \text{WebS} & 2.89 & 0.07 \\
 \text{C4} & \text{WebV} & 3.37 & 0.09 \\
 \text{C4} & \text{TVAd} & 4.68 & 0.11 \\
 \text{C4} & \text{P} & 6.12 & 0.11 \\
 \hline
 \text{C5} & \text{Q} & 2.88 & 0.22 \\
 \text{C5} & \text{P} & 6.01 & 0.33 \\
 \hline
 \text{D1} & \text{Q} & 5.24 & 0.09 \\
 \text{D1} & \text{P} & 5.47 & 0.11 \\
 \text{D1} & \text{TVAd} & 5.84 & 0.15 \\
 \text{D1} & \text{WebS} & 6.07 & 0.07 \\
 \hline
 \text{D2} & \text{MobV} & 3.05 & 0.06 \\
 \text{D2} & \text{WebV} & 3.54 & 0.07 \\
 \text{D2} & \text{WebS} & 5.04 & 0.08 \\
 \text{D2} & \text{P} & 5.52 & 0.09 \\
 \text{D2} & \text{Q} & 5.70 & 0.09 \\
 \hline
 \text{D3} & \text{TVAd} & 3.87 & 0.19 \\
 \text{D3} & \text{Q} & 4.77 & 0.07 \\
 \hline
      \end{tabular}}
    \end{center}
    \caption{-continued from \figref{tab:A1left}}
    \label{tab:A1right}
  \end{minipage}
\end{table}

\begin{table}[ht]
  \begin{minipage}[t]{.45\textwidth}
    \begin{center}
     \scalebox{0.7}{
      \begin{tabular}{c|c|cc}
\hline
Brand&Variable&Phase&Abs.\\
\hline
\text{A1} & \text{TVAd} & 5.49 & 0.10 \\
 \text{A1} & \text{Q} & 5.70 & 0.17 \\
 \text{A1} & \text{P} & 5.84 & 0.09 \\
 \hline
 \text{A2} & \text{TVAd} & 4.52 & 0.09 \\
 \text{A2} & \text{Q} & 5.73 & 0.11 \\
 \hline
 \text{A3} & \text{MobS} & 4.77 & 0.07 \\
 \text{A3} & \text{TVAd} & 5.2 & 0.08 \\
 \text{A3} & \text{Q} & 5.54 & 0.18 \\
 \text{A3} & \text{P} & 5.92 & 0.08 \\
 \text{A3} & \text{MobV} & 6.06 & 0.07 \\
 \hline
 \text{A4} & \text{WebS} & 4.09 & 0.08 \\
 \text{A4} & \text{TVAd} & 4.76 & 0.13 \\
 \text{A4} & \text{WebV} & 5.24 & 0.12 \\
 \text{A4} & \text{Q} & 5.80 & 0.20 \\
 \text{A4} & \text{P} & 6.11 & 0.14 \\
 \hline
 \text{A6} & \text{P} & 0.73 & 0.06 \\
 \hline
 \text{A7} & \text{Q} & 4.92 & 0.12 \\
 \hline
 \text{B1} & \text{P} & 5.19 & 0.10 \\
 \text{B1} & \text{WebS} & 5.28 & 0.07 \\
 \text{B1} & \text{Q} & 5.36 & 0.18 \\
 \text{B1} & \text{TVAd} & 5.53 & 0.08 \\
 \hline
 \text{B2} & \text{Q} & 4.06 & 0.16 \\
 \text{B2} & \text{WebS} & 5.02 & 0.07 \\
 \text{B2} & \text{P} & 5.27 & 0.09 \\
 \text{B2} & \text{TVAd} & 5.64 & 0.11 \\
 \hline
 \text{B3} & \text{WebV} & 2.91 & 0.08 \\
 \text{B3} & \text{TVAd} & 5.21 & 0.09 \\
 \hline
 \text{B4} & \text{Q} & 5.15 & 0.19 \\
 \text{B4} & \text{P} & 5.46 & 0.12 \\
  \hline
      \end{tabular}}
    \end{center}
    \caption{List of components in the second eigenmode with absolute value above the 2$\sigma$ range
    (to be continued to \figref{tab:A2right})}
    \label{tab:A2left}
  \end{minipage}
  %
  %
  \begin{minipage}[t]{.45\textwidth}
    \begin{center}
      \scalebox{0.7}{
      \begin{tabular}{c|c|cc}
\hline
Brand&Variable&Phase&Abs.\\
\hline
\text{B5} & \text{Q} & 5.07 & 0.11 \\
 \text{B5} & \text{P} & 5.36 & 0.10 \\
 \text{B5} & \text{TVAd} & 5.63 & 0.10 \\
 \hline
 \text{C1} & \text{WebV} & 2.02 & 0.07 \\
 \text{C1} & \text{TVAd} & 5.08 & 0.19 \\
 \text{C1} & \text{WebS} & 5.25 & 0.07 \\
 \text{C1} & \text{Q} & 5.96 & 0.12 \\
 \text{C1} & \text{P} & 6.28 & 0.08 \\
 \hline
 \text{C2} & \text{TVAd} & 5.22 & 0.12 \\
 \text{C2} & \text{P} & 5.37 & 0.08 \\
 \text{C2} & \text{Q} & 5.49 & 0.13 \\
 \hline
 \text{C3} & \text{Q} & 3.18 & 0.12 \\
 \text{C3} & \text{P} & 3.38 & 0.18 \\
 \text{C3} & \text{TVAd} & 3.69 & 0.12 \\
 \text{C3} & \text{WebV} & 5.2 & 0.08 \\
 \hline
 \text{C4} & \text{TVAd} & 4.47 & 0.09 \\
 \text{C4} & \text{WebV} & 5.14 & 0.07 \\
 \text{C4} & \text{Q} & 5.34 & 0.22 \\
 \hline
 \text{C5} & \text{P} & 3.10 & 0.18 \\
 \text{C5} & \text{Q} & 5.99 & 0.23 \\
 \hline
 \text{D1} & \text{Q} & 5.52 & 0.19 \\
 \text{D1} & \text{WebS} & 5.74 & 0.07 \\
 \text{D1} & \text{P} & 5.81 & 0.14 \\
 \hline
 \text{D2} & \text{WebS} & 5.34 & 0.08 \\
 \text{D2} & \text{Q} & 5.34 & 0.18 \\
 \text{D2} & \text{P} & 5.39 & 0.12 \\
 \hline
 \text{D2} & \text{TVAd} & 5.58 & 0.12 \\
 \text{D3} & \text{TVAd} & 0.29 & 0.12 \\
 \text{D3} & \text{P} & 5.14 & 0.08 \\
 \text{D3} & \text{Q} & 5.81 & 0.21 \\
 \hline
      \end{tabular}}
    \end{center}
    \caption{-continued from \figref{tab:A2left}}
    \label{tab:A2right}
  \end{minipage}
\end{table}

\end{document}